\documentclass[article, final, nonacm]{acmart} 
\usepackage{booktabs} 
\usepackage{wrapfig}

\begin{document}

\title{UrbanFlow: Designing Comfortable Outdoor Areas}
\author{Daoming Liu$^1$, Florian Rist$^1$, Helmut Pottmann$^1$, Dominik L.~Michels$^2$}
\email{{daoming.liu,florian.rist,helmut.pottmann,dominik.michels}@kaust.edu.sa}
\affiliation{
\institution{\\$^1$Computational Design and Fabrication Research Group,\\
$^2$Computational Sciences Group,
\\KAUST Visual Computing Center}
\country{Saudi Arabia}
}

\begin{abstract}
Design decisions in urban planning have to be made with particular carefulness as the resulting constraints are binding for the whole architectural design that follows. In this context, investigating and optimizing the airflow in urban environments is critical to design comfortable outdoor areas as unwanted effects such as windy areas and the formation of heat pockets have to be avoided. Our \textit{UrbanFlow} framework enables interactive architectural design allowing for decision making based on simulating urban flow. Compared to real-time fluid flow simulation, enabling interactive architecture design poses an even higher computational efficiency challenge as evaluating a design by simulation usually requires hundreds of time steps. This is addressed based on a highly efficient Eulerian fluid simulator in which we incorporate a unified porosity model which is devised to encode digital urban models containing objects such as buildings and trees. \textit{UrbanFlow} is equipped with an optimization routine enabling the direct computation of design adaptations improving livability and comfort for given parameterized architectural designs. To ensure convergence of the optimization process, instead of the classical Navier–Stokes equations, the Reynolds-averaged Navier–Stokes equations are solved as this can be done on a relatively coarse grid and allows for the decoupling of the effects of turbulent eddies which are taken into account using a separate turbulence model. As we demonstrate on a real-world example taken from an ongoing architectural competition, this results in a fast convergence of the optimization process which computes a design adaptation avoiding heat pockets as well as uncomfortable windy areas. \textit{UrbanFlow} exploits the power of the graphics processing unit running on a single desktop computer as it is widely available in most architectural and urban planning firms. We also provide a plugin which enable its use with the \textit{Rhinoceros 3D} software widely used in computational design and architecture.
\end{abstract}


\maketitle

\section{Introduction}
Urban planning deals with the design of settlements, building groups, city districts and public spaces in general.
It can be understood as an early stage of architectural design or its preceding process.
Design decisions have to be made with particular carefulness as the resulting constraints are binding for the whole architectural design that follows.
To avoid bad decision making, carefully investigating potential implications is absolutely critical.
In this regard, architects and urban planners have a vast interest in understanding and controlling the airflow within and around the particular design area as well as in its surrounding neighborhood.
This understanding can be used to reduce the energy consumption of a building in hot and cold climate~\cite{RN231, RN235}, improve the outdoor comfort~\cite{RN240}, and reduce building costs by controlling wind loads~\cite{RN238, RN239}.
Saving energy is of paramount importance since the building sector is currently responsible for about 40\% of the global carbon emissions~\cite{RN230}. Improving natural ventilation not only reduces the energy consumption of a building~\cite{RN232, RN236}, on a larger scale it also plays a key role in ensuring sufficient ventilation in urban areas~\cite{RN241, RN242} and creating a livable outdoor environment.
The effect of global warming is amplified in cities by the formation of heat pockets~\cite{RN229}.
Optimizing the airflow can help to mitigate this effect, and to create a livable and comfortable outdoor environment.

Air flow in an urban environment has to be considered as highly complex, and a designer can usually not rely on intuition to successfully utilize it. Computational fluid dynamics (CFD) can provide the necessary information to allow the designer to take informed decisions and improve the design. This information is already needed in early design phases, because fundamental decisions about the layout and morphology of the design are difficult to change in later design phases. Current simulation tools do not fit the specific needs of the architects in these early design phases. Compared to real-time fluid flow simulation, enabling interactive architecture design poses an even higher computational efficiency challenge as evaluating a design by simulation usually requires hundreds of time steps.

We present \textit{UrbanFlow}, an interactive simulation and optimization framework which can be integrated into the design workflow right from conceptual design on. It is seamlessly integrated into a computer-aided design (CAD) system often used by architects, allows for interactive design as well as automated optimizations. It further includes a porosity model as it is often overlooked that a city does not only consist of buildings, but important porous structures like trees as well. At the core of \textit{UrbanFlow} is our simulator designed as an Eulerian fluid solver. From a technical perspective, we had to address several points to enable the interactive flow simulation in urban environments. First, simply representing buildings as solid objects implies significant limitations as solid-wall boundary conditions for every object have to be enforced. Moreover, in an interactive design mode, the coefficient matrix for the pressure projection equation has to be reconstructed whenever the geometric configuration of a building is edited. As conjugate gradient (CG) preconditioners usually depend on the coefficient matrix, additional re-computations would be required substantially reducing the efficiency. Also, we would like to point out that some buildings are not purely solid objects dependent on their specific architecture potentially containing open design elements. Consequently, we propose to represent buildings as porous material instead of regular solid objects, and a high-level porosity model is constructed based on a widely-used empirical model for trees. For the urban flow simulation, an extra drag force term is used instead of enforcing the solid-wall boundary conditions. This way, the porosity effect of buildings can be easily included. More important, it can decouple the dependence between the coefficient matrix of the pressure projection equation of the CG preconditioner as well as the geometrical configuration of buildings. Therefore, the efficiency of the urban flow simulation can be improved using precomputation. However, although the CG preconditioner can be precomputed, its application process cannot be precomputed and may be quite time-consuming for some complicated preconditioners. This work explores an approximate inverse preconditioner that has not been widely used in graphics literature so far. Compared to other commonly used preconditioners, it shows superior balance between convergence rate and parallel-computing costs.

Our fluid solver supports arbitrary building shapes efficiently in order to facilitate a creative design process. In general, we need to know for each grid cell if it is empty or belongs to an object within the urban environment. Our building surfaces are geometrically represented as closed polygon meshes. Consequently, we adopt a classical even-odd rule which counts how many times a ray starting at the specific point crosses boundaries.The whole urban scene is discretized using a set of cells. For each cell we store a porosity value as a new unified way to encode urban models containing elements such as buildings and trees, as well as empty regions. Cut-cells can also be naturally handled by averaging the porosity values of their vertices. Motivated by this novel perspective, we further enable a new design pattern in which the porosity values are encoded by color and directly set by the designer. This way, the users could perform the design very easily in 2D, e.g., by painting. In parallel, the geometrical configurations of buildings and trees could be easily obtained by decoding the color of each grid cell without any explicit geometric calculations. Computational costs are further reduced and do not depend on the complexity of the building and tree shapes. This idea could also be easily extended to 3D for a given user-interaction interface which allows to directly access the 3D grid cells.

Next to enabling interactive architectural design, we further propose an optimization routine which directly computes slight design adaptations improving livability and comfort for given parameterized building layouts while maintaining the overall concept and structure of the design. The optimization routine employs a gradient descent procedure to compute the design adaptations which poses the challenges to avoid oscillations within the optimization process and to ensure convergence within a reasonable number of iterations. This is not easily possible when using the classical formulation of the Navier–Stokes (NS) equations due to the turbulent nature of the resulting flow. Instead, the Reynolds-averaged NS (RANS) equations allow for a decoupling of the effects of turbulent eddies which are taken into account using a separate turbulence model resulting in fast convergence of the optimization.

Our specific contributions are as follows. (1) A unified porosity model is devised to encode digital urban models containing objects such as buildings and trees. (2) An Eulerian urban flow simulator is presented showing high performance due to the pre-computation of the pressure projection equation's coefficient matrix and its CG preconditioner. (3) An approximate inverse preconditioner is explored and evaluated for the flow simulation showing excellent balance between convergence rate and computational costs. (4) An optimization procedure based on solving RANS equations is devised improving urban designs as demonstrated on a real-world example from an ongoing architectural competition. (5) We provide a plugin to couple our framework with the \textit{Rhinoceros 3D} (\textit{Rhino}) software which is widely used in computational design and architecture, and exploit the power of parallel hardware by implementing our framework on the graphics processing unit (GPU).

\section{Related Work}
In this section, we review related contributions.

\textit{Eulerian Fluid Simulation.}
At least since the seminal work of Stam~\shortcite{stam1999stable}, the simulation of fluids is an established research field within computer graphics. Back then, Stam proposed a semi-Lagrangian-type scheme to handle the nonlinear advection term within the NS equations based on Chorin's \shortcite{chorin1990mathematical} framework. Unconditionally stable fluid simulations could be achieved which are very attractive for graphics applications but suffer from severe numerical diffusion. Subsequent approaches have been proposed to overcome this limitation including the Mac-Cormack scheme \cite{selle2008unconditionally} which is adopted in our work, vorticity confinement~\cite{zhang2015restoring}, the advection-reflection scheme~\cite{zehnder2018advection}, and bidirectional mapping~\cite{qu2019efficient}. It is also worthy to mention that it has been claimed \cite{10.5555/3056883.3056894} that a full NS-based model would be too complicated to efficiently compute urban flow for which reason a light weight Lattice Boltzmann method has been employed for 2D urban flow simulations. In contrast, our work devises a fast NS equation solver for 3D urban flow.

The most time-consuming operation within the numerical integration of the NS equations is usually the pressure projection step in which a large scale pressure Poisson equation set has to be solved. Consequently, significant effort has been invested to accelerate this step. In general, iterative solvers are much more preferred than direct methods, since they behave much better in terms of computation time and memory costs for such large scale linear systems. Among many iterative methods, Krylov subspace approaches~\cite{saad2003iterative} such as preconditioned conjugate gradient (PCG) methods are widely adopted in graphics literature as the coefficient matrix of the pressure Poisson equations is usually symmetric and positive definite. The convergence rate of such PCG methods depends heavily on the efficacy of their preconditioners. The list of popular preconditioners include Jacobi, symmetric successive over-relaxation (SSOR), incomplete Cholesky factorization~\cite{foster2001practical, selle2005vortex}, as well as multigrid methods~\cite{mcadams2010parallel,aanjaneya2019efficient}. In our work, a kind of an approximate inverse preconditioner not widely used before in graphics is explored and evaluated against these popular preconditioners.

\textit{Optimizing Designs based on Fluid Simulation.}
Among others, shape optimization based on fluid simulation is an important topic with several potential applications. Plenty of work has been conducted in the CFD community, especially for the automobile and airplane industries, and became recently popular for closed indoor environment research~\cite{liu2015state}. Topology optimization also shares some similarities, and we refer to the review paper of Deaton and Grandhi~\shortcite{deaton2014survey} for more details. Within the graphics community, Du et al.~\shortcite{du2020functional} recently worked on design optimization for fluidic devices considering Stokes flow. This can be treated as the first step in this direction as claimed in their paper. Our work will target at the more challenging turbulent flow and the architecture design application in particular.

\textit{Computational Architectural Design.}
This field is of interest to many researchers within the graphics and CAD communities. The textbook of Pottmann et al.~\shortcite{pottmann2008architectural} provides an introduction to the underlying concepts and applications. For more recent advances in this field we refer to a survey~\cite{pottmann2015architectural} and a recent publication~\cite{gavriil2020computational}. However, most of this prior contributions were focused on the form/shape design in contemporary architecture. We further explore the optimization of designs by incorporating physical simulation, and in particular the air flow simulation within complex urban models. 

\section{Fast Urban Flow Simulation}
In this section, we introduce the technical details behind the fluid simulator within our \textit{UrbanFlow} framework.

\textit{Incompressible Navier-Stokes Equations.} The incompressive NS equations are given by
\begin{equation}
   \nabla \cdot \mathbf{u'} = 0\,,
   \label{eq:incom}
\end{equation}
\begin{equation}
		\frac{\partial \mathbf{u'}}{\partial t} + ( \mathbf{u'} \cdot \nabla )\mathbf{u'} = -\frac{1}{\rho_\mathsf{air}}\nabla p' + \nu \nabla^2 \mathbf{u'}\,,
\label{eq:momen}
\end{equation}
in which $\mathbf{u'}$ denotes the velocity, $p'$ the pressure, $\nu$ the kinematic viscosity of the air, and $\rho_\mathsf{air}$ the density of the air. The zero divergence condition in Eq.~(\ref{eq:incom}) ensures incompressibility of the air flow. Eq.~(\ref{eq:momen}) is also called momentum equation.

\textit{Reynolds-averaged Navier–Stokes Equations.} In the industrial design practice, especially for turbulent flow, the incompressible NS equations were rarely solved in a direct fashion as this requires a high grid resolution to capture the smallest eddy scale (i.e., Kolmogorov scale, around $Re^{3/4}$ for Reynolds number $Re$). Instead, the Reynolds-averaged NS (RANS) equations are widely used and implemented as a mainstream method in commercial CFD software~\cite{manceau2021industrial}. RANS equations can be solved on a relative coarse grid with the effect of turbulent eddy modelled with an additional turbulence model. Moreover, in our optimization routine (see Section~\ref{sec:opt}) we observe that the directly solved incompressible NS model can produce serious osculations not achieving convergence. RANS is a physical approximation of the original NS model derived via a Reynolds decomposition, Reynolds averaging and the Boussinesq relation. The mathematical formulations of the time-dependent RANS \cite{hanjalic2020eddy} are given by Eq.~(\ref{eq:incom}) and
\begin{equation}
  	\frac{\partial \mathbf{u}}{\partial t} + ( \mathbf{u} \cdot \nabla )\mathbf{u} = -\frac{1}{\rho_\mathsf{air}}\nabla p + (\nu+\nu_t) \nabla^2 \mathbf{u} + \frac{1}{\rho_\text{air}}\mathbf{f_d}\,,
		\label{eq:momen_drag}
\end{equation}
\begin{equation}
    \mathbf{u} =\mathbf{u'}-\mathbf{\tilde{u}}, ~~~p=p'-\tilde{p}\,,
\end{equation}
in which $(\mathbf{u}, p)$ denotes the mean (i.e., time-averaged) part of $(\mathbf{u'}, p')$, and $(\mathbf{\tilde{u}}, \tilde{p})$ denotes the fluctuating part according to the Reynolds decomposition.
Motivated by the widely used drag force model for trees~\cite{kenjerevs2013modelling,kang2017development}, we have included a drag force $\mathbf{f_d} = - \rho_\mathsf{air}~C_d~\text{LAD}~|\mathbf{u}|~\mathbf{u}$, in which $C_d$ denotes the drag coefficient which depends on the leaf roughness, the leaf area density $LAD$ denotes the total one-side leaf area per unit volume, $\mathbf{u}$ denotes the wind velocity and $|\mathbf{u}|$ its magnitude.\footnote{A high-level drag force model for buildings can be constructed if LAD is replaced with the term $a~((1 - \phi)/(\phi + e))^b$. Here, $\phi$ denotes the porosity coefficient of the building and $e=1^{-10}$ is a small number to avoid zero-division. The coefficients $a=0.62$ and $b=2.5$ have been experimentally validated.}

\textit{Turbulence Model.} Intuitively, the eddy viscosity term ($\nu_t \gg \nu$) in the RANS model is used to smooth the mean flow. This is why it has been called ``laminarization'' in the original paper~\cite{jones1972prediction} of the popular $k-\varepsilon$ turbulence model which considers the turbulent kinetic energy $k$ and the turbulent kinetic energy dissipation rate $\varepsilon$. We have extensively tested this model in our numerical experiments and observed serious numerical instability problems. Hence, we further implemented its alternative variant, the $k-\omega$ turbulence model \cite{wilcox2006turbulence} (small-eddy frequency $\omega=\varepsilon / (C_{\mu}k)$ with a model parameter $C_{\mu}$) given by
\begin{equation}
    \frac{\partial k}{\partial t} + (\mathbf{u}\cdot \nabla) k = P_k - C_{\mu} k\omega + (\nu + \sigma^{*} \nu_t)\nabla^2 k\,,
\end{equation}
\begin{equation}
    \frac{\partial \omega}{\partial t} + (\mathbf{u}\cdot \nabla) \omega = 2\alpha |S|^2 - \beta \omega^2 + (\nu + \sigma \nu_t)\nabla^2 \omega\,,
\end{equation}
\begin{equation}
    P_k=2\nu_t|S|^2\,,~~~~~~S = (\nabla \mathbf{u} + \left(\nabla \mathbf{u}\right)^\mathsf{T})/2\,,
\end{equation}
\begin{equation}
\nu_t = \frac{k}{\tilde{\omega}},~~~\tilde{\omega} = \max\{\omega,~ C_\text{lim}|S|/(C_{\mu}/2)\}\,.
\end{equation}
The mean strain-rate tensor is denoted by $S$. The values of $k$ and $\omega$ on the inlet boundary are calculated as $k=1.5\,(I\,U_\text{ref})^2$ and $\omega = {L}^{-1}{C_{\mu}^{-0.25}k^{0.5}}$, in which $I$ denotes the turbulence intensity, and $U_\text{ref}$ and $L$ denote the reference speed and length scale respectively.
The model's coefficients are shown in Table~\ref{table:tc}.
\begin{table}[htb]
	\centering 
	\caption{Coefficients within the $k-\omega$ model.} \label{table:tc}  
	\begin{tabular*}{0.65\columnwidth}{@{}@{\extracolsep{\fill}}cccccc@{}}
			\hline  
			 $C_\mu$ & $\alpha$ & $\beta$ & $\sigma$ & $\sigma^{*}$ & $C_\text{lim}$ \\
			\hline  
			0.09 & 0.52 & 0.0708 & 0.5 & 0.6 & 7/8\\   
			\hline  
	\end{tabular*} 	  
\end{table}

\textit{Discretization and Numerical Integration.} While the $k-\omega$ model shows improved numerical robustness, it still limits the use of a larger time step (around 1/10) compared to the case without using a decoupled turbulence model. To support a stable simulation with a larger time step, we propose the following viscosity limiter, which is derived via the von Neumann stability analysis on the diffusion term of Eq.~(\ref{eq:momen_drag}):
\begin{equation}
    \tilde{\nu_t} = \min\left\{\nu_t, ~~{2\Delta t ({\Delta x}^2 + {\Delta y}^2 + {\Delta z}^2)^{-1} } - \nu\right\}\,.
\end{equation}
In \textit{UrbanFlow}, the time-dependent RANS (or called U/T RANS) equations share the same mathematical form with NS except an extra eddy-viscosity parameter, then can be solve with fast simulation methods developed in computer graphics literature. In this work, we adopt the Eulerian Fluid simulation framework with a uniform staggered Cartesian grid. The components $(u, v, w)$ of the velocity field $\mathbf{u}$ are located at the face center of each grid cell while pressure $p$ and external drag force $\mathbf{f_d}$ stored at the center of each grid cell. A time splitting method is used to handle the time derivative term of the equation, and all other terms are temporally integrated one by one. The Mac-Cormack-based semi-Lagrangian scheme is adopted to handle the nonlinear advection term, and the pressure Poisson equation is solved by the conjugate gradient algorithm (CG) with an approximate inverse (AI) preconditioner. The additional turbulence model equations are discretized with a forward difference scheme for the time derivative term, central difference scheme for the Laplacian diffusion term, and upwind scheme for the advection term. Please note, that the turbulence equations have been solved after the projection step so that the incompressive velocity data can be used.

\textit{Approximate Inverse Preconditioner.}
GPU parallelization is exploited to boost up \textit{UrbanFlow}'s performance. One of the most important parts is to handle the preconditioned CG algorithm that is adopted to iteratively solve the large scale pressure Poisson equation in each time step. However, the modified incomplete Cholesky decomposition preconditioner (widely used in serial code) cannot be easily parallelized, because it involves forward and backward substitutions to solve the triangular matrix equation which are inherently serial procedures~\cite{chu2017schur}. The algebraic and geometric multigrid preconditioners are too complicated for our application. The simple Jacobi preconditioner is straightforward for the parallel implementation on the GPU, but it only comes with slight improvements on the convergence rate compared to the naive CG algorithm. A kind of preconditioner that was not widely used before in graphics literature is explored in \textit{UrbanFlow}. It was firstly proposed by Ament at al.~\shortcite{ament2010parallel} and was called incomplete Poisson (IP) preconditioner in their original article. Later on, a slightly modified formulation was rigorously derived from the first order Neumann series approximation of the SSOR preconditioner by Helfenstein and Koko~\shortcite{helfenstein2012parallel}. When solving a system $\mathbf{A}\mathbf{x}=\mathbf{b}$, the basic idea of such a preconditioner is to find a direct approximating matrix $\mathbf{M}^{-1}$ of $\mathbf{A}^{-1}$, instead of the approximating matrix $\mathbf{M}$ of $\mathbf{A}$. It turned out that it can be applied as simple as the Jacobi preconditioner through one matrix-vector multiplication while almost equally good results compared to other more advanced preconditioners. This is very attractive for our work. Its formulation is as follows:
\begin{equation}
    \mathbf{M}^{-1} = \mathbf{K}^\mathsf{T}\mathbf{K}\,,
\end{equation}
in which the first-order and second-order approximations of $K$ are
\begin{equation}
    \mathbf{K} = \sqrt{2-\omega}\bar{\mathbf{D}}^{-1/2}(\mathbf{I} - \mathbf{L} \bar{\mathbf{D}}^{-1})\,,
\end{equation}
\begin{equation}
    \mathbf{K} = \sqrt{2-\omega}\bar{\mathbf{D}}^{-1/2}(\mathbf{I} - \mathbf{L} \bar{\mathbf{D}}^{-1} + (\mathbf{L} \bar{\mathbf{D}}^{-1})^2)\,,
\end{equation}
where $0<\omega<2$, $\bar{\mathbf{D}} = \mathbf{D}/\omega$, and $\mathbf{L}$ and $\mathbf{D}$ denote the lower triangle part and the diagonal part of the matrix $\mathbf{A}$, respectively, which can be easily calculated. The incomplete formulation of the matrix $\mathbf{M}^{-1}$ can be calculated by keeping the same sparsity as in the matrix $\mathbf{A}$. In our implementation, all these sparse matrices are stored in the compressed sparse row (CSR) format. In addition, the user's interactive editing of the urban model is decoupled from $\mathbf{A}$, and thus all these calculations of the preconditioner matrix $\mathbf{M^{-1}}$ can be precomputed for our simulation. Even in other applications where it cannot be precomputed, it still can be efficiently calculated since it is straightforward to be parallelized on the GPU.

\textit{Porosity Model vs.~Solid-wall Boundary Conditions.}
\label{sec:bc}
Setting up appropriate boundary conditions is critical for the fluid simulation and can be very tricky. In this work, we use the Dirichlet condition to directly set the velocity for the inlet boundary, while for the outlet boundary we use the Neumann boundary condition to ensure that the derivative of the velocity (with respect to the normal direction of the outlet boundary) is equal to zero:
\begin{equation}
\partial \mathbf{u} / \partial \mathbf{n} = 0\,.
\end{equation}
No-slip solid wall boundary conditions are enforced over the solid wall boundary:
\begin{equation}
    \mathbf{u} = \mathbf{u}_\text{solid}\,.
\end{equation}
In the pressure projection step, the pressure in the outlet cell is set to be zero, and a ghost pressure value in the inlet and solid wall cell is calculated by a Neumann boundary condition on the fly:
\begin{equation}
\partial p / \partial\mathbf{n} = 0\,.
\label{eq:pressure}
\end{equation}

We would like to particularly discuss more about the handling of the boundary condition of buildings which is very important in our application. If we treat it as a solid wall, besides the fact that its porosity effect cannot be included, extra inefficiency will also be caused by the solid-wall boundary condition enforced over the surfaces of the buildings. As described above, the wall boundary condition needs to be specially handled, and the geometry configuration of buildings will be constantly changed for interactive urban design. Moreover, the coefficient matrix $\mathbf{A}$ of the pressure Poisson equation set depends on the buildings' geometrical configuration since a ghost pressure value needs be calculated for the solid cells nearby the solid-wall boundary according to Eq.~(\ref{eq:pressure}). It turned out that each time the user edits the urban model, the coefficient matrix $\mathbf{A}$ and its preconditioner have to be recomputed. When the porosity model is used to handle buildings, an extra drag force will be enforced for each building cell and no solid-wall boundary conditions are needed anymore. Since the coefficient matrix $\mathbf{A}$ does not depend on the drag force term, this problem is solved. The coefficient matrix $\mathbf{A}$ and its preconditioner can be precomputed to improve the computational efficiency.

We use the forward Euler method to numerically solve the drag force step:
\begin{equation}
    \frac{\partial \mathbf{u}}{\partial t} = \frac{1}{\rho_\mathsf{air}} \mathbf{f_d} =
    - C_d a( (1-\phi)/(\phi + e))^b U \mathbf{u}\,,
\end{equation}
\begin{equation}
    \mathbf{u}_{n+1} = (1- \Delta t C_d a( (1-\phi)/(\phi + e))^b U) \mathbf{u}_n\,,
\end{equation}
in which the constraint $(1- \Delta t C_d a ((1 - \phi)/(\phi + e))^b U) \geq 0$ is enforced and any negative values will be replaced by zeros. It is clear that this drag force term damps the velocity field and for the zero-porosity case (i.e., solid cell), the speed is always ensured to be zero.  

\textit{Urban model resolving.} To efficiently handle the buildings, trees, and the boundary setting of the simulation domain, we use the marked-cell method to assign a label value to each grid cell to show its type. The possible types for boundary cells include inlet, outlet, and solid wall, which can be easily set. The possible types for internal cells include buildings, trees and air, while its setting will take some effort because we need to resolve the urban model first and the urban model will be constantly edited by the designers in the interactive mode or keep evolving in the automatic optimization mode. 

A two-pass strategy is adopted. For a given geometry object (building or tree) with arbitrary shape (its surface is represented as a triangle mesh), an axis-align bounding box (AABB) is generated in the first pass. It can be easily judged if some grid cell $\mathbf{c}$ is inside its AABB or not (we use cell center points in our implementation for simplicity). If grid cell $\mathbf{c}$ is inside its AABB, we use the ray-casting algorithm and the odd-even rule to determine if this grid cell is inside this object or not in the second pass. The ray is used to calculate its intersection points with all the surface mesh triangles of this building if any. We count the number of intersection points, and an odd number means that this grid cell is inside this object. The intersection of ray and triangle is calculated with the classical Möller–Trumbore algorithm~\shortcite{moller1997fast}.

\begin{figure}[htb]
    \centering
    \vspace{-0.7cm}
    \includegraphics[width=0.35\columnwidth]{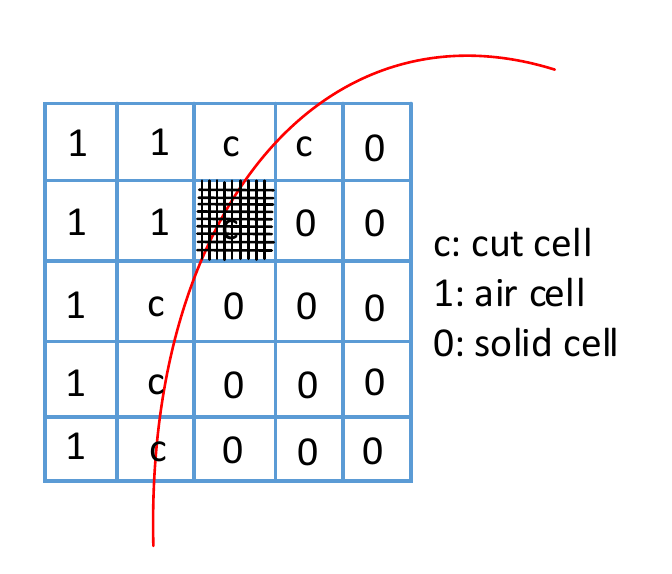}
    \vspace{-0.35cm}
    \caption{Cut-cell treatment: The red curve denotes the solid/building boundary, porosity of 0 denotes the cell inside and porosity of 1 denotes the cell outside the object. The cut cell is refined to get a high sub-cell resolution to more accurately resolve the boundary. The volume ration is calculated as its porosity, which is very important for the sliding boundary within the cell in the optimization iteration.}
    \label{fig:cut-cell}
\end{figure}

The grid cells can be classified into three types: Inside-architecture cells, outside-architecture cells, and the cut-cells by the surfaces of the architectures. Particular treatment needs to be done with the cut-cells to better support sub-cell resolution for architecture surfaces, continuous optimization computation, simplicity and efficiency \cite{du2020functional}. In this paper, we propose to use the volume ratio outside architecture surface as the porosity values of the cut-cells. Different from the plane fitting method using least square algorithm based on the signed distance calculated on the quadrature points in cut-cells adopted from Du et al.~\shortcite{du2020functional}, we adopt an alternative method where a uniform subdivision to the desired resolution is performed in the cut-cells, and the ratio of the subdivision grid points outside the architectures is used as the volume ratio approximation instead of the linear plane approximation. This is illustrated in Figure~\ref{fig:cut-cell}. 

\section{Towards Simulation-inspired Design}

\subsection{Interactive Design}
Because of the fact that most of the urban designers are not so familiar with fluid simulation, an interface tool with popular urban design software needs to be developed where the designers can easily run the flow simulation in the way that they are familiar with.

\textit{Rhino Plugin.} We developed a plugin to use \textit{UrbanFlow} with \textit{Rhino} widely used in computational architecture. This supports designers to perform the interactive design of urban models directly in a software system they are familiar with. An initial urban model can be automatically generated from the procedural urban model plugin or just a non-procedural urban model can be imported, and then designers perform edits. Whenever the designer would like to know about the flow distribution under some geometrical configuration, the urban flow simulation can be called via the developed urban flow plugin. Then, the geometry information of the urban model will be transferred to our fast simulator, and the simulation results are transferred back to \textit{Rhino} for visualization/rendering. Besides the urban flow plugin, a simple procedural urban model plugin has been developed that can produce random urban models for testing.

\subsection{Automatic Design Optimization}
\label{sec:opt}
In this section, we discuss the details of \textit{UrbanFlow}'s optimization routine.

\textit{Objective Function.} Optimizing air flow is the main focus of this work and several objective functions have been defined for different specific applications. Generally, the objective function is defined as
\begin{equation}
    L(\mathbf{u}) = \sum_i||F_i(\mathbf{u}) - F_i(\mathbf{u_t})||_2,~~\mathbf{u} = \Phi(\theta)\,,
\end{equation}
in which $\mathbf{u_t}$ denotes the desired velocity field and $F_i$ denotes a function of the velocity field, $\Phi$ denotes the forward simulation operator, and $\theta$ denotes the design parameter. The index $i$ corresponds to a specific target region.

\textit{Gradient-descent.} The standard gradient-descent optimization algorithm is implemented in this work. One key issue is the computation of the gradient of objective functions with respect to the design parameters. A simple strategy is used in this paper, where the well-designed fast forward simulation is used to calculate the gradient information. Its mathematical formulation is given by
\begin{equation}
    \mathbf{\theta}^*:=\mathsf{argmin}_{\mathbf{\theta}}\{L(\mathbf{\theta})\}\,,
\end{equation}
\begin{equation}
    \mathbf{\theta}^{n+1} = \mathbf{\theta}^{n} + \lambda \frac{\partial L}{\partial \mathbf{\theta}},
    ~~~~~~~~~~~~~~~~~~~~\frac{\partial L}{\partial \theta_i} \approx \frac{L(\theta_i + \epsilon ) - L(\theta_i)}{\epsilon}\,,
    \label{eqn:gradienopti}
\end{equation}
in which $\mathbf{\theta}$ denotes a vector with $N$ design parameters $\theta_i$, $L$ denotes the objective function, and $\lambda$ denotes the step size. The partial derivative with respect to the design parameter $\theta_i$ is calculated by forward differences, where a small increment $\epsilon$ is added to the specific parameter. In addition, the wind direction perturbation is an important factor that needs to be taken into account especially for outdoor ventilation.

\textit{Discussion.} This work is mainly targeted at the early stage urban design application where buildings are usually represented with simple shapes and small number of design parameters are needed. Therefore, we adopt the light-weight numerical method based on forward simulation to calculate gradient instead of the more complicated advanced methods such as adjoint method, auto-differentiation, etc. These advanced methods will take extra cost by themselves and thus are more suitable for the applications with large number of design parameters.

\section{Implementation}
\begin{figure}[htb]
    \centering
    \includegraphics[width=0.99\columnwidth]{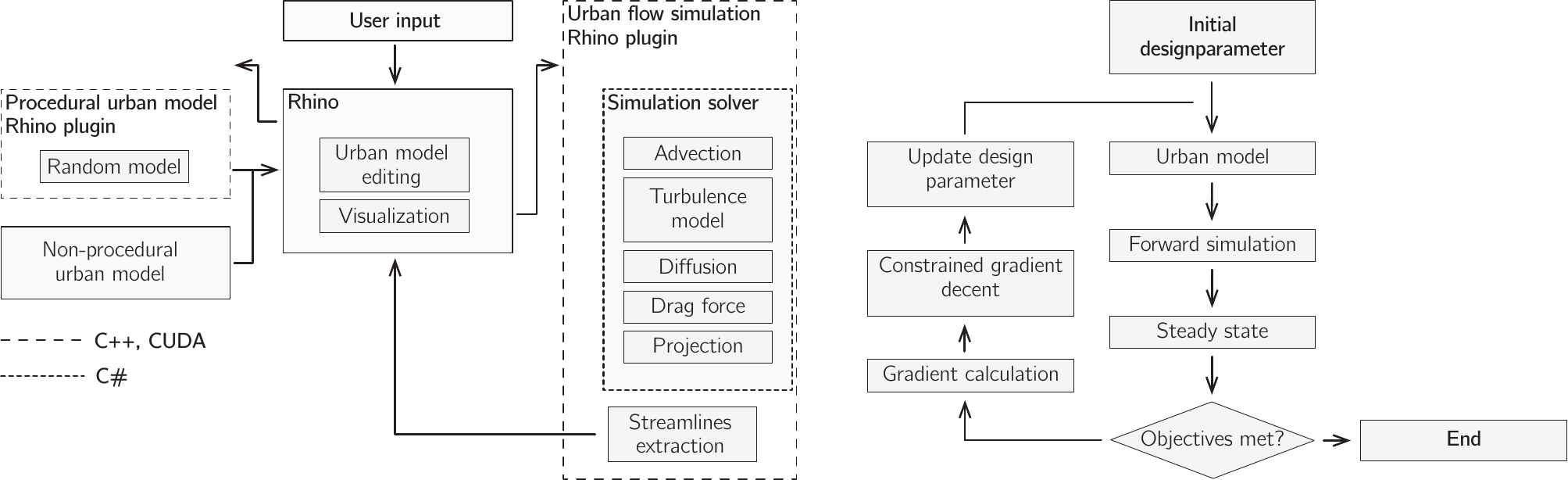}
    \caption{System diagram of the \textit{Rhino} plugin developed for interactive design (left); flow chart of the optimization procedure (right).}
    \label{fig:system_diagram}
\end{figure}

All results presented in this paper are computed on a workstation with an Intel Xeon CPU E5-2699, an NVIDIA TITAN V GPU, and 256 GB main memory. The flow simulation solver, urban model handling, and the generation and extraction of the streamlines are implemented in C++ and parallelized using CUDA. In the interactive mode, they are compiled into a dynamic link library (DLL) that is called by the \textit{Rhino} interfaces/plugins which are implemented in C$\#$ and \textit{RinoCommon}. The calculation of the matrix condition number is done with \textit{MATLAB}. The optimization routine is also implemented with our in-housed code in C++ and CUDA, and no pre-existing packages are used. The visualization/rendering is mainly done with \textit{Rhino}. Figure \ref{fig:system_diagram} (left) shows the diagram of the Rhino plugin developed for the interactive design. The procedure of the optimization scheme is presented in Figure \ref{fig:system_diagram} (right).

\section{Case studies for Urban design applications}
\label{sec:res}
We have carefully validated our fluid simulator including its porosity model and the choice of the AI preconditioner. This can be found in the appendix.

\begin{figure*}[t]
\centering
\includegraphics[width=0.33\textwidth]{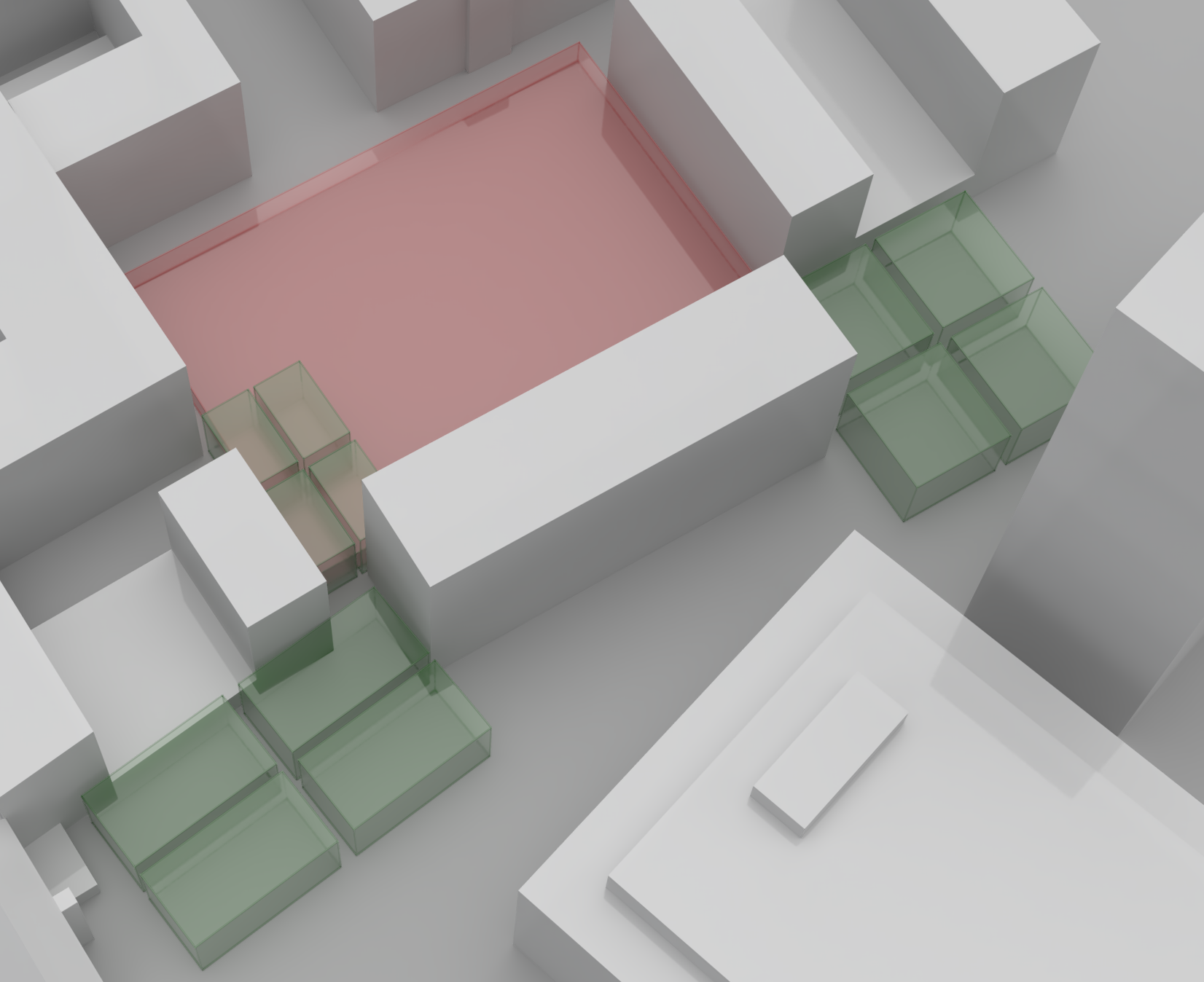}
\includegraphics[width=0.33\textwidth]{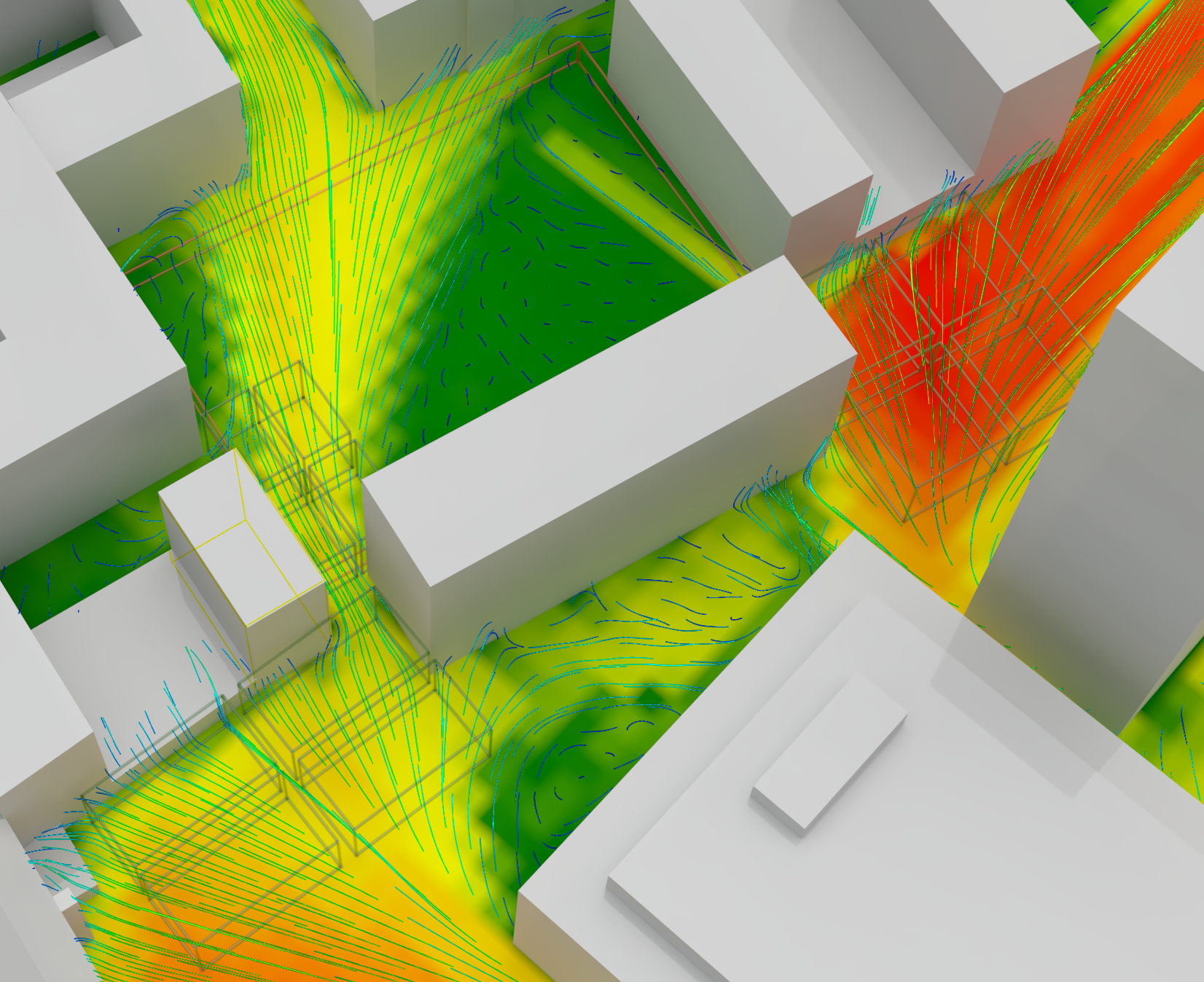}
\includegraphics[width=0.33\textwidth]{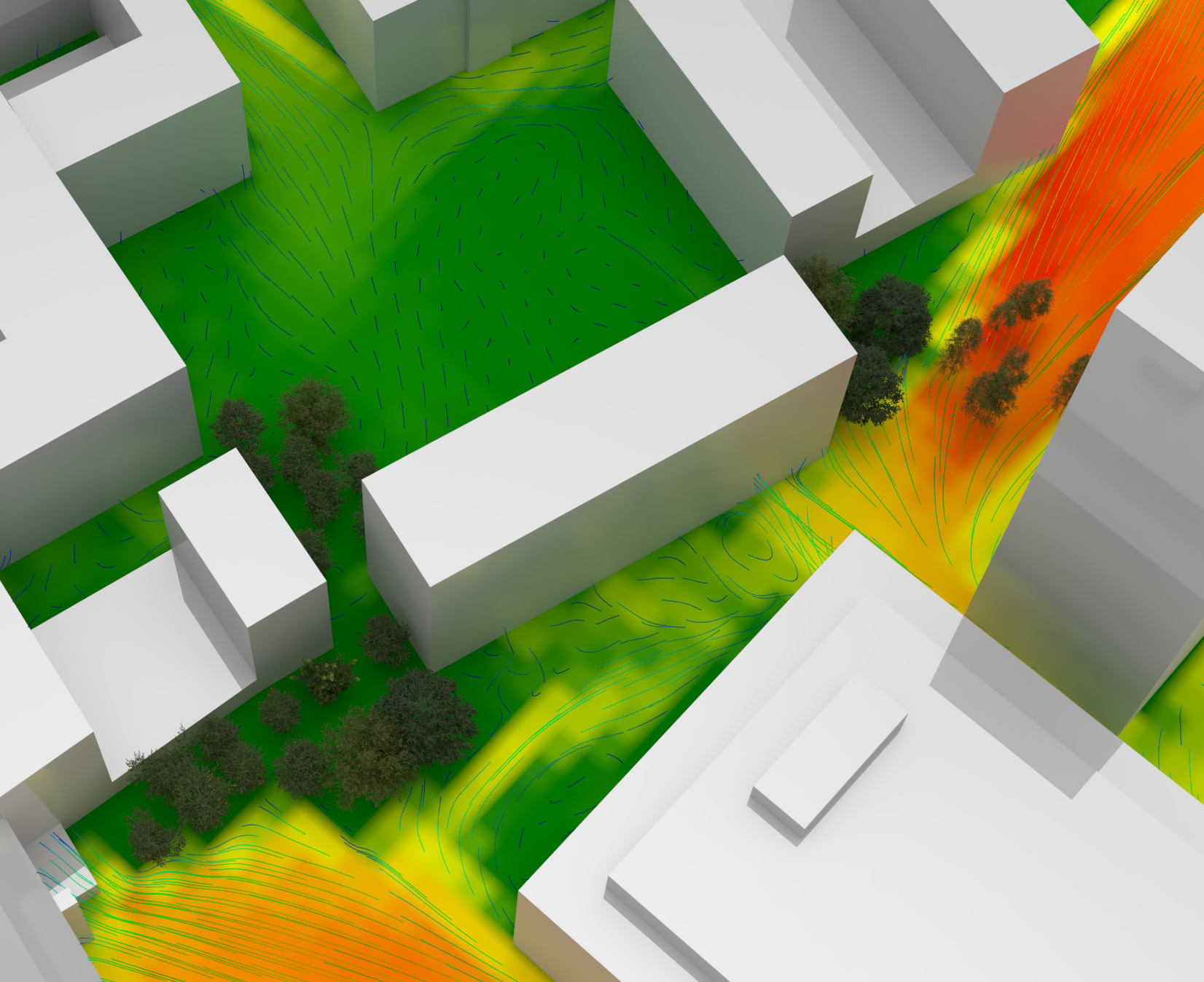}\\
\vspace{-0.3cm}
\caption{Simulation of the urban flow at the windy public ``Neumarkt'' square in Bielefeld, Germany (middle). The outdoor comfort could be improved by planting trees in strategic locations which are determined using our system. The area of interest (red) and possible plantation locations (green) are shown (left). A possible landscaping following the optimized porosity is illustrated (right). }
\label{fig:teaser}
\end{figure*}

\subsection{Improving Wind Comfort by Trees Planting}

Figure~\ref{fig:teaser} (left) shows the digital urban model of the ``Neumarkt'' square in the city of Bielefeld in Germany. This square is known as an extremely windy place within the city as strong winds are coming from southwest (bottom left in our illustrations). First, we run an urban flow simulation using real wind data obtained from the Global Wind Atlas\footnote{http://globalwindatlas.info/}. A uniform grid with the resolution of $100 \times 100 \times 30$ is used and each grid cell contains a volume of $3.5 \times 3.5 \times 3.5$~m$^3$. The outflow boundary condition is enforced over right and top boundaries. Over the inflow (left and bottom boundaries), a vertical wind inflow with a logarithmic profile $u_z = {u_*}/{\kappa}\ln\left({z}/{z_0}\right)$ is enforced, in which $u_z$ denotes the speed at altitude $z$, $\kappa\approx0.41$ the dimensionless Von K\'arm\'an constant, $z_0$ the roughness length of 0.5~m, and $u_*=0.53$~ms$^{-1}$ the friction speed calculated by data-fitting with real meteorological data of Bielefeld.

\begin{figure}[htb]
   \centering
    \includegraphics[width=0.25\columnwidth]{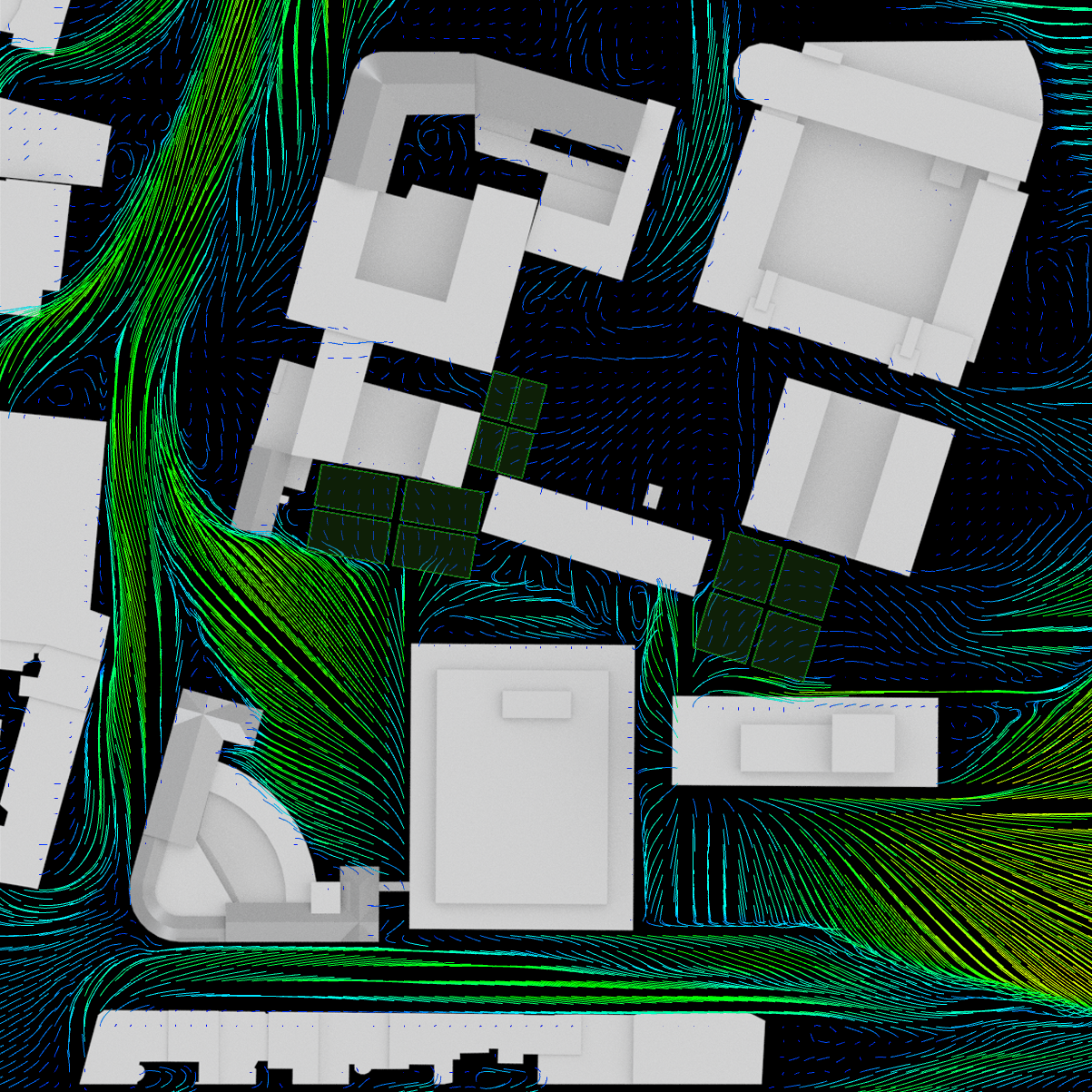}
    \includegraphics[width=0.25\columnwidth]{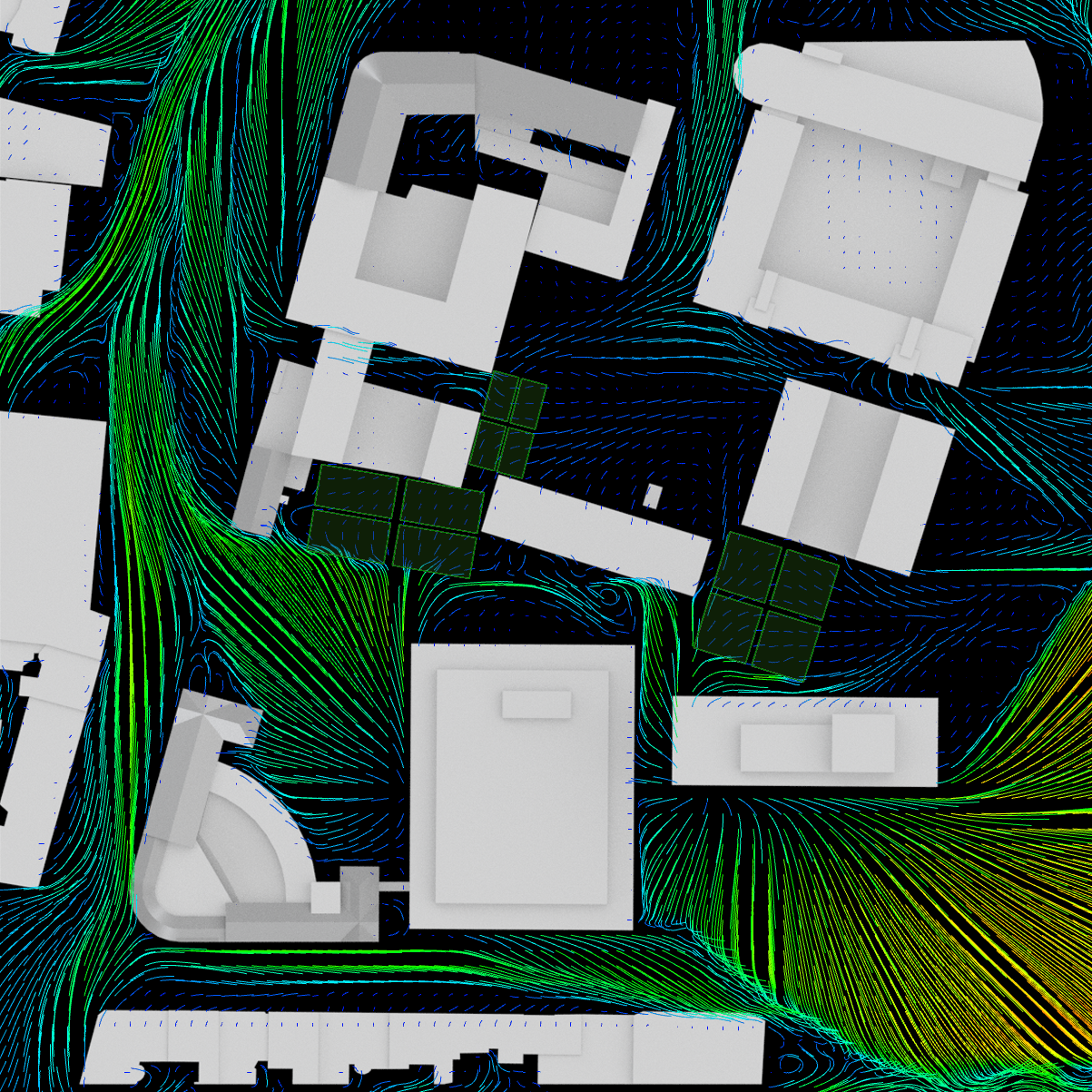}\\
    \vspace{-0.35cm}
    \caption{Test of the tree planting solution against perturbations of the main wind direction ($30^\circ$). Simulations using wind directions of $20^\circ$ (left) and $40^\circ$ are shown.}
    \label{fig:tree_plant_robust}
\end{figure}

As shown in Figure~\ref{fig:teaser} (middle), uncomfortable wind speeds are present on the square. A natural question is if we can overcome this issue by planting trees. After some interactive operations shown in our supplemental video, we have been able to identify a design solution of planting trees that can clearly improve the windy square problem. The urban flow simulation results for the improved situation after planting trees is shown in Figure~\ref{fig:teaser} (right). To test the robustness of this design solution under perturbations of the main wind direction, two more simulations with $\pm 10^\circ$ wind direction perturbations have been carried out as shown in Figure~\ref{fig:tree_plant_robust}. It can be observed that the right-bottom corner of the square can always be well improved, while the left-top corner is more complicated.

\begin{figure*}
    \centering
    \includegraphics[width=1.0\textwidth]{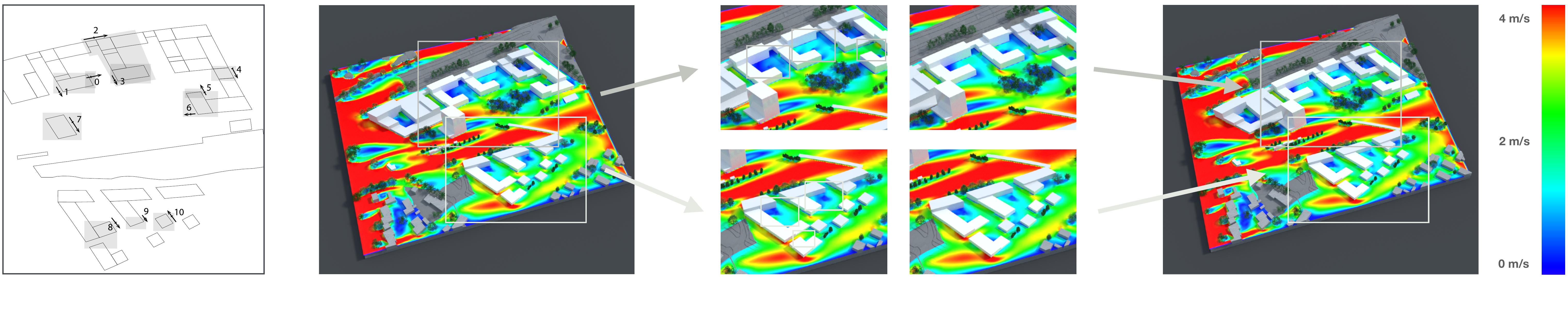}\\
    \vspace{-0.6cm}
    \caption{Illustration of \textit{UrbanFlow}’s optimization process: After the identification of DOFs by a professional architect familiar with the initial design (left), a forward simulation is carried out to identify target areas in which the wind speed should be improved (second and third images from the left). After the gradient descent method converges, which computes the design adaptations, the effect of the new design is shown (fourth and fifth image from the left). It avoids heat pockets without compromising the overall outdoor comfort. The wind speed is illustrated using a color map. Please note, that low wind speeds (highlighted in blue) correspond to potential heat pockets. 
    }
    \label{fig:Nurnberg_optimization1}
\end{figure*}

\begin{figure}[htb]
   \centering
    \includegraphics[height=0.29\columnwidth]{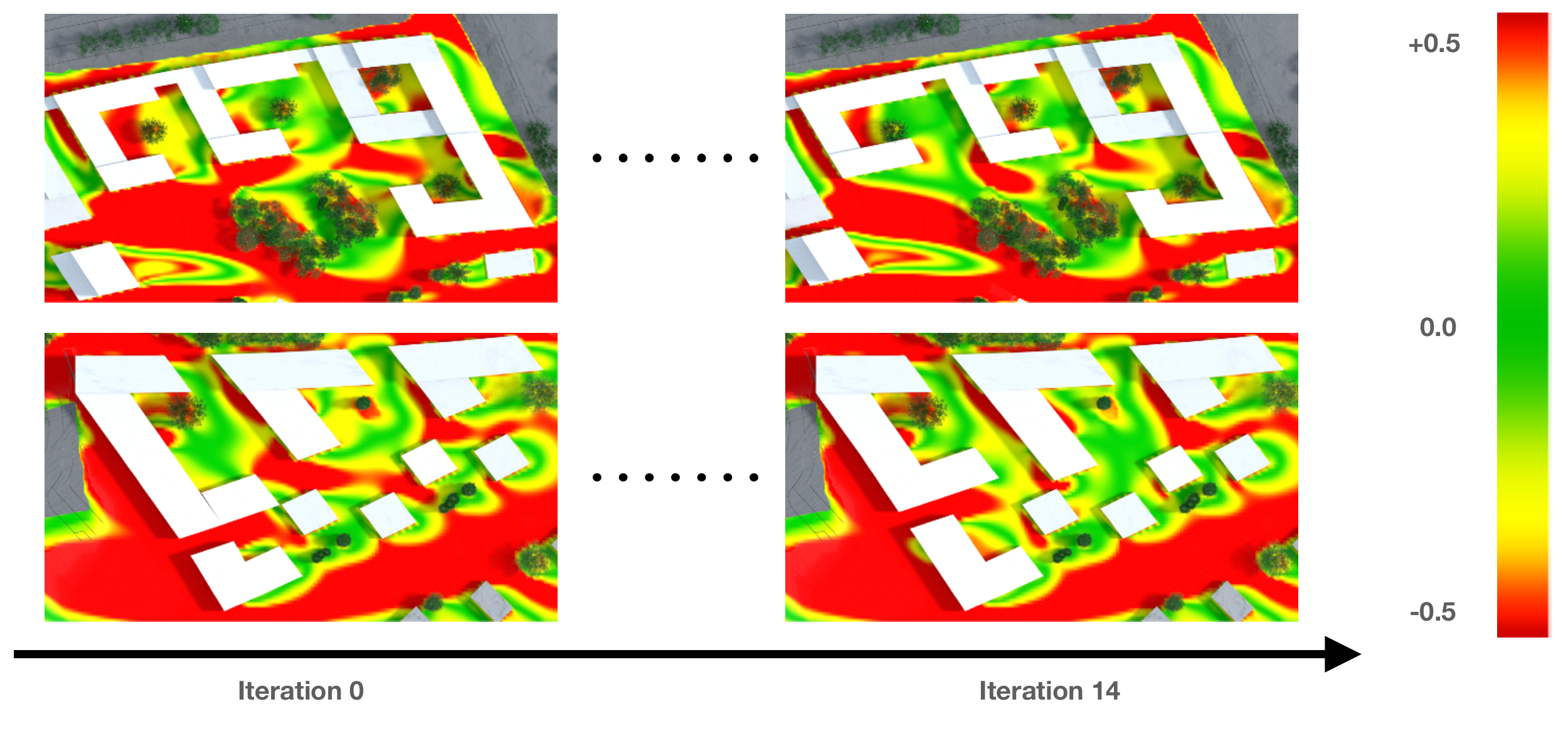}
    \hspace{2mm}
    \includegraphics[height=0.29\columnwidth]{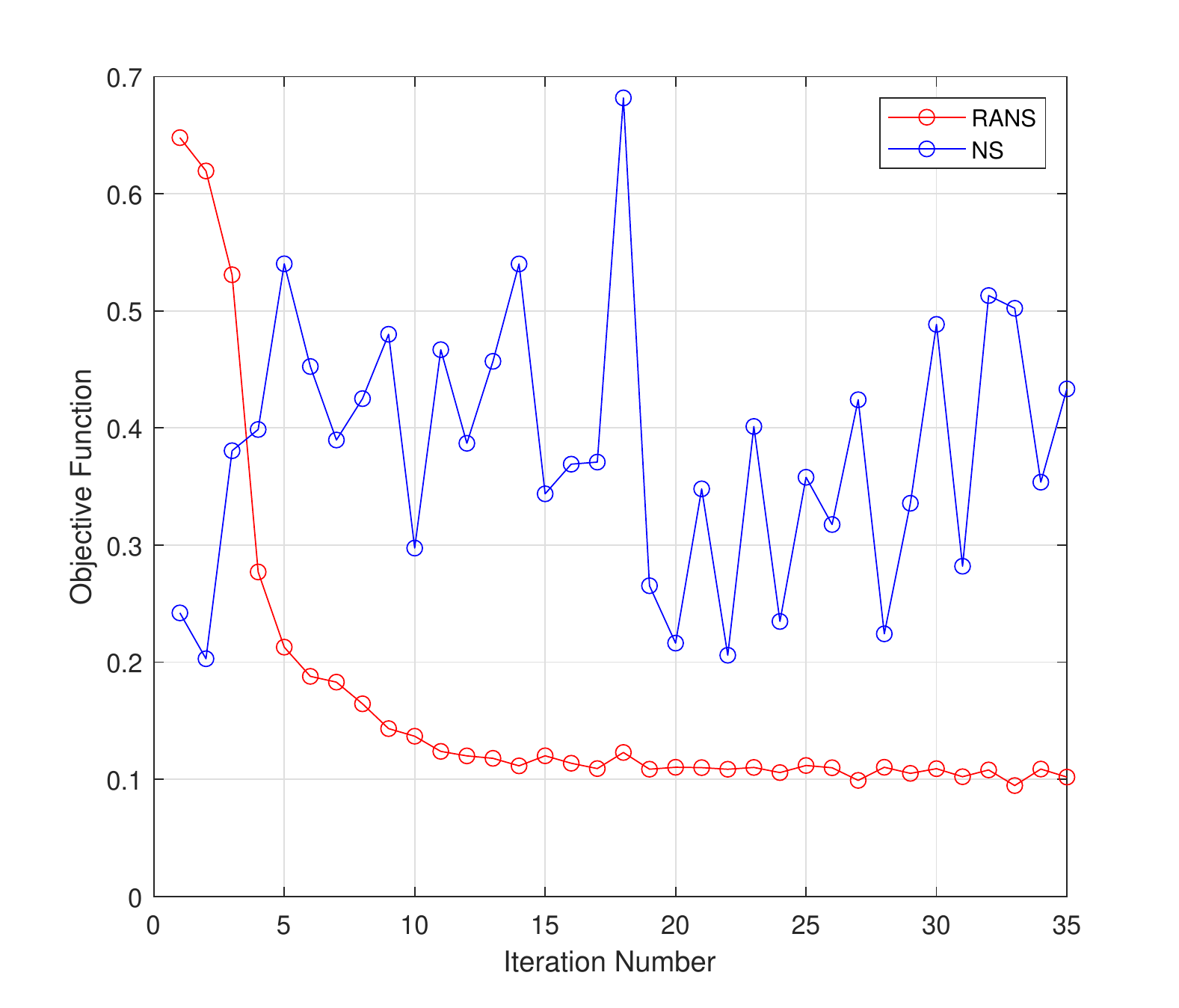}\\
    \vspace{-0.4cm}
    \caption{Illustration of the result of the optimization routine (left) and the development of the objective function during the optimization process (right). The deviation from the target wind speed of 0.55~m/s is illustrated using a color map.}
    \label{fig:Nurnberg_optimization2}
\end{figure}

\vspace{-0.2cm}
\subsection{Avoiding Heat Pockets in Early Design Stages}
We demonstrate the practical usability of \textit{UrbanFlow}’s optimization routine by contributing to an ongoing architectural competition based on an urban design by \textit{BLAUWERK}, a well established architectural firm in Munich specialized on residential and office buildings, and \textit{bauchplan}, a landscape architectural firm mostly working on designing public spaces. Recently, these firms have jointly worked on a design for a new residential quarter on a 44\,000~${\text{m}}^2$ area in Nürnberg-Gebersdorf in Germany. One of the design objectives given to the architects has been to prevent the formation of heat pockets in high density areas with low ventilation. The particular challenge is to compute small design adaptations in order to increase outdoor ventilation in a way that the formation of heat pockets is avoided while at the same time no uncomfortable windy areas should be created. In order to keep the spirit and the visual impression of the design\footnote{Please note, that in January 2022 this design draft was awarded a third price in an international competition organized by the developer \textit{WBG Nürnberg GmbH} in cooperation with the \textit{Bavarian State Ministry of Housing, Building and Transport} and the \textit{Bavarian State Ministry for the Environment and Consumer Protection} underlying how important it is to not change the visual appearance too much.}, and do not change its key properties too much (e.g., the overall amount of office space), only small adaptations should be made.

The design is shown in Figure~\ref{fig:Nurnberg_optimization1} in which the critical areas where heat pockets are expected are surrounded with rectangular frames (second image from the left). An architect who is familiar with this design systematically identified eleven degrees of freedom (DOFs) which can be easily changed in the design draft. These DOFs are shown in Figure~\ref{fig:Nurnberg_optimization1} (left). Moreover, maximum values for the adaptation of all DOFs have been defined to ensure physical constraints so that, e.g., buildings cannot collide or overlap. An objective function
\begin{equation}
L(\tilde{\bf{u}}) = \sum_{i=1}^N (\tilde{u_i}-u_t)^2
\label{eqn:targetfunction}
\end{equation}
has been defined in which $N=6$ denotes the number of target regions under consideration in which heat pockets or inconvenient windy area can occur.\footnote{Please note, that among these six target regions we do not only consider regions in which heat pockets can occur but also regions which are potentially exposed to inconveniently high wind speeds. This is especially important in this application as strong wind blow from the west side because there is mostly undeveloped land.} For illustration, the six target regions are surrounded with rectangular frames in Figure~\ref{fig:Nurnberg_optimization1} (middle). In Eq.~(\ref{eqn:targetfunction}), $\tilde{\bf{u}}$ denotes the vector of all average wind speeds $\tilde{u_i}$ (each corresponding to the $i$-th target region) and $u_t$ denotes the ideal average wind speed which we set to 0.55~m/s which is sufficient to avoid heat pockets without compromising outdoor comfort. Driven by this objective function, the average wind speed should automatically approach the ideal wind speed $u_t$. However, please note that the average wind speed of some regions can naturally not converge to the ideal wind speed. Hence, the value of the objective function does not have to be zero when converged. In each iteration of the optimization process, the wind field is computed using a forward simulation using a uniform grid resolution of $450\times 300 \times 40$ and a cell volume of $1 \times 1 \times 1$~$\text{m}^3$. A time step of 0.2~s is applied for these forward simulation and about 2\,000 time steps are computed per forward simulation. The gradient descent algorithm then applies an optimization step size of $\lambda = 1.0$ and the small increment is set to $\epsilon=0.1$.

The optimization process is illustrated in Figure~\ref{fig:Nurnberg_optimization2} which shows that the gradient descent optimizer converges after 14 iterations using the RANS solver. According to Eq.~(\ref{eqn:gradienopti}), we have to compute $12$ (i.e.,number of parameters plus one) forward simulations in each iteration of the gradient descent routine. This results in a computation time of about 2~hours for all 14 iterations on a regular single workstation. In Figure~\ref{fig:Nurnberg_optimization2} (right), we have also shown what happens if we use the standard NS formulation instead of RANS which only results in oscillations not leading to convergence of the optimization routine. The overall result of the optimization process is shown in Figure~\ref{fig:Nurnberg_optimization1} demonstrating a significant improvement of the initial design with just minor adaptations necessary. The formation of heat pockets is avoided where possible without compromising outdoor comfort and maintaining the nature of the award winning design. After obtaining the design solution, forward simulations with small perturbations of the main wind direction are carried out in order to verify the robustness of the result. 

\section{Discussion, Limitations, and Future Work}
In this contribution, we have demonstrated how fluid flow simulation can contribute to intelligent decision making in architectural design. This is based on our \textit{UrbanFlow} framework which combines an efficient Eulerian fluid simulator with a unified porosity model to encode digital urban models containing objects such as buildings and trees. \textit{UrbanFlow} has been used in the context of two real-world design challenges improving the wind comfort on a square in Bielefeld and for avoiding heat pockets in a current project in Nürnberg. While the wind comfort in the first application has been improved by planting trees based on design decisions obtained by running our flow simulator, the optimization routine of \textit{UrbanFlow} has been applied in the second application resulting in the computation of a design adaptation without a human in the loop. These practical examples showcase the usefulness of simulation-inspired design decisions in architecture and urban planning. While the optimization routine keeps the overall structure on the initial design, from a technical perspective, key objective of the design such as the amount of office space, are not strictly enforced as hard constraints. Overcoming this limitation by extending our work with a constraint solver is left for future work. In general, we consider \textit{UrbanFlow} as a first step towards the integration of physics-simulations in the design process of architects opening up an interesting avenue of future work including applications such as designing public areas in a way that noise in minimized.

\bibliographystyle{ACM-Reference-Format}
\bibliography{acmart}


\begin{thebibliography}{39}


\ifx \showCODEN    \undefined \def \showCODEN     #1{\unskip}     \fi
\ifx \showDOI      \undefined \def \showDOI       #1{#1}\fi
\ifx \showISBNx    \undefined \def \showISBNx     #1{\unskip}     \fi
\ifx \showISBNxiii \undefined \def \showISBNxiii  #1{\unskip}     \fi
\ifx \showISSN     \undefined \def \showISSN      #1{\unskip}     \fi
\ifx \showLCCN     \undefined \def \showLCCN      #1{\unskip}     \fi
\ifx \shownote     \undefined \def \shownote      #1{#1}          \fi
\ifx \showarticletitle \undefined \def \showarticletitle #1{#1}   \fi
\ifx \showURL      \undefined \def \showURL       {\relax}        \fi
\providecommand\bibfield[2]{#2}
\providecommand\bibinfo[2]{#2}
\providecommand\natexlab[1]{#1}
\providecommand\showeprint[2][]{arXiv:#2}

\bibitem[\protect\citeauthoryear{Aanjaneya, Han, Goldade, and Batty}{Aanjaneya
  et~al\mbox{.}}{2019}]%
        {aanjaneya2019efficient}
\bibfield{author}{\bibinfo{person}{Mridul Aanjaneya},
  \bibinfo{person}{Chengguizi Han}, \bibinfo{person}{Ryan Goldade}, {and}
  \bibinfo{person}{Christopher Batty}.} \bibinfo{year}{2019}\natexlab{}.
\newblock \showarticletitle{An Efficient Geometric Multigrid Solver for Viscous
  Liquids}.
\newblock \bibinfo{journal}{\emph{Proceedings of the ACM on Computer Graphics
  and Interactive Techniques}} \bibinfo{volume}{2}, \bibinfo{number}{2}
  (\bibinfo{year}{2019}), \bibinfo{pages}{1--21}.
\newblock


\bibitem[\protect\citeauthoryear{Adamek, Vasan, Elshaer, English, and
  Bitsuamlak}{Adamek et~al\mbox{.}}{2017}]%
        {RN240}
\bibfield{author}{\bibinfo{person}{Kimberley Adamek}, \bibinfo{person}{Neetha
  Vasan}, \bibinfo{person}{Ahmed Elshaer}, \bibinfo{person}{Elizabeth English},
  {and} \bibinfo{person}{Girma Bitsuamlak}.} \bibinfo{year}{2017}\natexlab{}.
\newblock \showarticletitle{Pedestrian level wind assessment through city
  development: A study of the financial district in Toronto}.
\newblock \bibinfo{journal}{\emph{Sustainable cities and society}}
  \bibinfo{volume}{35} (\bibinfo{year}{2017}), \bibinfo{pages}{178--190}.
\newblock
\showISSN{2210-6707}


\bibitem[\protect\citeauthoryear{Al-Saadi and Shaaban}{Al-Saadi and
  Shaaban}{2019}]%
        {RN235}
\bibfield{author}{\bibinfo{person}{Saleh~Nasser Al-Saadi} {and}
  \bibinfo{person}{Awni~K Shaaban}.} \bibinfo{year}{2019}\natexlab{}.
\newblock \showarticletitle{Zero energy building (ZEB) in a cooling dominated
  climate of Oman: Design and energy performance analysis}.
\newblock \bibinfo{journal}{\emph{Renewable and Sustainable Energy Reviews}}
  \bibinfo{volume}{112} (\bibinfo{year}{2019}), \bibinfo{pages}{299--316}.
\newblock
\showISSN{1364-0321}


\bibitem[\protect\citeauthoryear{Ament, Knittel, Weiskopf, and Strasser}{Ament
  et~al\mbox{.}}{2010}]%
        {ament2010parallel}
\bibfield{author}{\bibinfo{person}{Marco Ament}, \bibinfo{person}{Gunter
  Knittel}, \bibinfo{person}{Daniel Weiskopf}, {and} \bibinfo{person}{Wolfgang
  Strasser}.} \bibinfo{year}{2010}\natexlab{}.
\newblock \showarticletitle{A parallel preconditioned conjugate gradient solver
  for the Poisson problem on a multi-GPU platform}. In
  \bibinfo{booktitle}{\emph{2010 18th Euromicro Conference on Parallel,
  Distributed and Network-based Processing}}. IEEE, \bibinfo{pages}{583--592}.
\newblock


\bibitem[\protect\citeauthoryear{Chen, Tong, and Malkawi}{Chen
  et~al\mbox{.}}{2017}]%
        {RN232}
\bibfield{author}{\bibinfo{person}{Yujiao Chen}, \bibinfo{person}{Zheming
  Tong}, {and} \bibinfo{person}{Ali Malkawi}.} \bibinfo{year}{2017}\natexlab{}.
\newblock \showarticletitle{Investigating natural ventilation potentials across
  the globe: Regional and climatic variations}.
\newblock \bibinfo{journal}{\emph{Building and Environment}}
  \bibinfo{volume}{122} (\bibinfo{year}{2017}), \bibinfo{pages}{386--396}.
\newblock
\showISSN{0360-1323}


\bibitem[\protect\citeauthoryear{Chorin, Marsden, and Marsden}{Chorin
  et~al\mbox{.}}{1990}]%
        {chorin1990mathematical}
\bibfield{author}{\bibinfo{person}{Alexandre~Joel Chorin},
  \bibinfo{person}{Jerrold~E Marsden}, {and} \bibinfo{person}{Jerrold~E
  Marsden}.} \bibinfo{year}{1990}\natexlab{}.
\newblock \bibinfo{booktitle}{\emph{A mathematical introduction to fluid
  mechanics}}. Vol.~\bibinfo{volume}{168}.
\newblock \bibinfo{publisher}{Springer}.
\newblock


\bibitem[\protect\citeauthoryear{Chu, Zafar, and Yang}{Chu
  et~al\mbox{.}}{2017}]%
        {chu2017schur}
\bibfield{author}{\bibinfo{person}{Jieyu Chu}, \bibinfo{person}{Nafees~Bin
  Zafar}, {and} \bibinfo{person}{Xubo Yang}.} \bibinfo{year}{2017}\natexlab{}.
\newblock \showarticletitle{A Schur complement preconditioner for scalable
  parallel fluid simulation}.
\newblock \bibinfo{journal}{\emph{ACM Transactions on Graphics (TOG)}}
  \bibinfo{volume}{36}, \bibinfo{number}{4} (\bibinfo{year}{2017}),
  \bibinfo{pages}{1}.
\newblock


\bibitem[\protect\citeauthoryear{Deaton and Grandhi}{Deaton and
  Grandhi}{2014}]%
        {deaton2014survey}
\bibfield{author}{\bibinfo{person}{Joshua~D Deaton} {and}
  \bibinfo{person}{Ramana~V Grandhi}.} \bibinfo{year}{2014}\natexlab{}.
\newblock \showarticletitle{A survey of structural and multidisciplinary
  continuum topology optimization: post 2000}.
\newblock \bibinfo{journal}{\emph{Structural and Multidisciplinary
  Optimization}} \bibinfo{volume}{49}, \bibinfo{number}{1}
  (\bibinfo{year}{2014}), \bibinfo{pages}{1--38}.
\newblock


\bibitem[\protect\citeauthoryear{Du, Wu, Spielberg, Matusik, Zhu, and
  Sifakis}{Du et~al\mbox{.}}{2020}]%
        {du2020functional}
\bibfield{author}{\bibinfo{person}{Tao Du}, \bibinfo{person}{Kui Wu},
  \bibinfo{person}{Andrew Spielberg}, \bibinfo{person}{Wojciech Matusik},
  \bibinfo{person}{Bo Zhu}, {and} \bibinfo{person}{Eftychios Sifakis}.}
  \bibinfo{year}{2020}\natexlab{}.
\newblock \showarticletitle{Functional optimization of fluidic devices with
  differentiable stokes flow}.
\newblock \bibinfo{journal}{\emph{ACM Transactions on Graphics (TOG)}}
  \bibinfo{volume}{39}, \bibinfo{number}{6} (\bibinfo{year}{2020}),
  \bibinfo{pages}{1--15}.
\newblock


\bibitem[\protect\citeauthoryear{Feng, Zhang, Ji, Wang, Zhou, Ye, Hao, Li, Luo,
  and Lau}{Feng et~al\mbox{.}}{2019}]%
        {RN231}
\bibfield{author}{\bibinfo{person}{Wei Feng}, \bibinfo{person}{Qianning Zhang},
  \bibinfo{person}{Hui Ji}, \bibinfo{person}{Ran Wang}, \bibinfo{person}{Nan
  Zhou}, \bibinfo{person}{Qing Ye}, \bibinfo{person}{Bin Hao},
  \bibinfo{person}{Yutong Li}, \bibinfo{person}{Duo Luo}, {and}
  \bibinfo{person}{Stephen Siu~Yu Lau}.} \bibinfo{year}{2019}\natexlab{}.
\newblock \showarticletitle{A review of net zero energy buildings in hot and
  humid climates: Experience learned from 34 case study buildings}.
\newblock \bibinfo{journal}{\emph{Renewable and Sustainable Energy Reviews}}
  \bibinfo{volume}{114} (\bibinfo{year}{2019}), \bibinfo{pages}{109303}.
\newblock
\showISSN{1364-0321}


\bibitem[\protect\citeauthoryear{Foster and Fedkiw}{Foster and Fedkiw}{2001}]%
        {foster2001practical}
\bibfield{author}{\bibinfo{person}{Nick Foster} {and} \bibinfo{person}{Ronald
  Fedkiw}.} \bibinfo{year}{2001}\natexlab{}.
\newblock \showarticletitle{Practical animation of liquids}. In
  \bibinfo{booktitle}{\emph{Proceedings of the 28th annual conference on
  Computer graphics and interactive techniques}}. \bibinfo{pages}{23--30}.
\newblock


\bibitem[\protect\citeauthoryear{Gavriil, Guseinov, P{\'e}rez, Pellis,
  Henderson, Rist, Pottmann, and Bickel}{Gavriil et~al\mbox{.}}{2020}]%
        {gavriil2020computational}
\bibfield{author}{\bibinfo{person}{Konstantinos Gavriil},
  \bibinfo{person}{Ruslan Guseinov}, \bibinfo{person}{Jes{\'u}s P{\'e}rez},
  \bibinfo{person}{Davide Pellis}, \bibinfo{person}{Paul Henderson},
  \bibinfo{person}{Florian Rist}, \bibinfo{person}{Helmut Pottmann}, {and}
  \bibinfo{person}{Bernd Bickel}.} \bibinfo{year}{2020}\natexlab{}.
\newblock \showarticletitle{Computational design of cold bent glass
  fa{\c{c}}ades}.
\newblock \bibinfo{journal}{\emph{ACM Transactions on Graphics (TOG)}}
  \bibinfo{volume}{39}, \bibinfo{number}{6} (\bibinfo{year}{2020}),
  \bibinfo{pages}{1--16}.
\newblock


\bibitem[\protect\citeauthoryear{Hanjali{\'c} and Launder}{Hanjali{\'c} and
  Launder}{2020}]%
        {hanjalic2020eddy}
\bibfield{author}{\bibinfo{person}{K Hanjali{\'c}} {and} \bibinfo{person}{BE
  Launder}.} \bibinfo{year}{2020}\natexlab{}.
\newblock \showarticletitle{Eddy-Viscosity Transport Modelling: A Historical
  Review}.
\newblock In \bibinfo{booktitle}{\emph{50 Years of CFD in Engineering
  Sciences}}. \bibinfo{publisher}{Springer}, \bibinfo{address}{Berlin},
  \bibinfo{pages}{295--316}.
\newblock


\bibitem[\protect\citeauthoryear{Helfenstein and Koko}{Helfenstein and
  Koko}{2012}]%
        {helfenstein2012parallel}
\bibfield{author}{\bibinfo{person}{Rudi Helfenstein} {and}
  \bibinfo{person}{Jonas Koko}.} \bibinfo{year}{2012}\natexlab{}.
\newblock \showarticletitle{Parallel preconditioned conjugate gradient
  algorithm on GPU}.
\newblock \bibinfo{journal}{\emph{J. Comput. Appl. Math.}}
  \bibinfo{volume}{236}, \bibinfo{number}{15} (\bibinfo{year}{2012}),
  \bibinfo{pages}{3584--3590}.
\newblock


\bibitem[\protect\citeauthoryear{Irving and Clements-Croome}{Irving and
  Clements-Croome}{2005}]%
        {RN236}
\bibfield{author}{\bibinfo{person}{Sam Irving} {and} \bibinfo{person}{Derek~J.
  Clements-Croome}.} \bibinfo{year}{2005}\natexlab{}.
\newblock \bibinfo{booktitle}{\emph{Natural ventilation in non-domestic
  buildings}}.
\newblock \bibinfo{publisher}{Chartered Institution of Building Services
  Engineers}.
\newblock
\showISBNx{0900953772}


\bibitem[\protect\citeauthoryear{Jones and Launder}{Jones and Launder}{1972}]%
        {jones1972prediction}
\bibfield{author}{\bibinfo{person}{W.~Peter Jones} {and}
  \bibinfo{person}{Brian~Edward Launder}.} \bibinfo{year}{1972}\natexlab{}.
\newblock \showarticletitle{The prediction of laminarization with a
  two-equation model of turbulence}.
\newblock \bibinfo{journal}{\emph{International journal of heat and mass
  transfer}} \bibinfo{volume}{15}, \bibinfo{number}{2} (\bibinfo{year}{1972}),
  \bibinfo{pages}{301--314}.
\newblock


\bibitem[\protect\citeauthoryear{Kang, Kim, Kim, Choi, and Park}{Kang
  et~al\mbox{.}}{2017}]%
        {kang2017development}
\bibfield{author}{\bibinfo{person}{Geon Kang}, \bibinfo{person}{Jae-Jin Kim},
  \bibinfo{person}{Dong-Ju Kim}, \bibinfo{person}{Wonsik Choi}, {and}
  \bibinfo{person}{Soo-Jin Park}.} \bibinfo{year}{2017}\natexlab{}.
\newblock \showarticletitle{Development of a computational fluid dynamics model
  with tree drag parameterizations: Application to pedestrian wind comfort in
  an urban area}.
\newblock \bibinfo{journal}{\emph{Building and Environment}}
  \bibinfo{volume}{124} (\bibinfo{year}{2017}), \bibinfo{pages}{209--218}.
\newblock


\bibitem[\protect\citeauthoryear{Kaseb, Hafezi, Tahbaz, and Delfani}{Kaseb
  et~al\mbox{.}}{2020}]%
        {RN241}
\bibfield{author}{\bibinfo{person}{Zeynab Kaseb}, \bibinfo{person}{Mohammadreza
  Hafezi}, \bibinfo{person}{Mansoureh Tahbaz}, {and} \bibinfo{person}{Shahram
  Delfani}.} \bibinfo{year}{2020}\natexlab{}.
\newblock \showarticletitle{A framework for pedestrian-level wind conditions
  improvement in urban areas: CFD simulation and optimization}.
\newblock \bibinfo{journal}{\emph{Building and Environment}}
  \bibinfo{volume}{184} (\bibinfo{year}{2020}), \bibinfo{pages}{107191}.
\newblock
\showISSN{0360-1323}


\bibitem[\protect\citeauthoryear{Kenjere{\v{s}} and ter Kuile}{Kenjere{\v{s}}
  and ter Kuile}{2013}]%
        {kenjerevs2013modelling}
\bibfield{author}{\bibinfo{person}{Sa{\v{s}}a Kenjere{\v{s}}} {and}
  \bibinfo{person}{Benjamin ter Kuile}.} \bibinfo{year}{2013}\natexlab{}.
\newblock \showarticletitle{Modelling and simulations of turbulent flows in
  urban areas with vegetation}.
\newblock \bibinfo{journal}{\emph{Journal of Wind Engineering and Industrial
  Aerodynamics}}  \bibinfo{volume}{123} (\bibinfo{year}{2013}),
  \bibinfo{pages}{43--55}.
\newblock


\bibitem[\protect\citeauthoryear{Liu, Zhang, Xue, Zhai, Wang, Wei, and
  Chen}{Liu et~al\mbox{.}}{2015}]%
        {liu2015state}
\bibfield{author}{\bibinfo{person}{Wei Liu}, \bibinfo{person}{Tengfei Zhang},
  \bibinfo{person}{Yu Xue}, \bibinfo{person}{Zhiqiang~John Zhai},
  \bibinfo{person}{Jihong Wang}, \bibinfo{person}{Yun Wei}, {and}
  \bibinfo{person}{Qingyan Chen}.} \bibinfo{year}{2015}\natexlab{}.
\newblock \showarticletitle{State-of-the-art methods for inverse design of an
  enclosed environment}.
\newblock \bibinfo{journal}{\emph{Building and Environment}}
  \bibinfo{volume}{91} (\bibinfo{year}{2015}), \bibinfo{pages}{91--100}.
\newblock


\bibitem[\protect\citeauthoryear{Lucon, \"Urge-Vorsatz, Ahmed, Akbari,
  Bertoldi, Cabeza, Eyre, Gadgil, Harvey, and Jiang}{Lucon
  et~al\mbox{.}}{2014}]%
        {RN230}
\bibfield{author}{\bibinfo{person}{Oswaldo Lucon}, \bibinfo{person}{Diana
  \"Urge-Vorsatz}, \bibinfo{person}{A~Zain Ahmed}, \bibinfo{person}{Hashem
  Akbari}, \bibinfo{person}{Paolo Bertoldi}, \bibinfo{person}{Luisa~F Cabeza},
  \bibinfo{person}{Nicholas Eyre}, \bibinfo{person}{Ashok Gadgil},
  \bibinfo{person}{L~Danny Harvey}, {and} \bibinfo{person}{Yi Jiang}.}
  \bibinfo{year}{2014}\natexlab{}.
\newblock \showarticletitle{Buildings}.
\newblock   \bibinfo{volume}{IPCC Fifth Assessment Report (AR5)}
  (\bibinfo{year}{2014}).
\newblock


\bibitem[\protect\citeauthoryear{Manceau}{Manceau}{2021}]%
        {manceau2021industrial}
\bibfield{author}{\bibinfo{person}{Remi Manceau}.}
  \bibinfo{year}{2021}\natexlab{}.
\newblock \bibinfo{booktitle}{\emph{Industrial codes for CFD}}.
\newblock \bibinfo{address}{Poitiers, France}.
\newblock


\bibitem[\protect\citeauthoryear{McAdams, Sifakis, and Teran}{McAdams
  et~al\mbox{.}}{2010}]%
        {mcadams2010parallel}
\bibfield{author}{\bibinfo{person}{Aleka McAdams}, \bibinfo{person}{Eftychios
  Sifakis}, {and} \bibinfo{person}{Joseph Teran}.}
  \bibinfo{year}{2010}\natexlab{}.
\newblock \showarticletitle{A Parallel Multigrid Poisson Solver for Fluids
  Simulation on Large Grids.}. In \bibinfo{booktitle}{\emph{Symposium on
  Computer Animation}}. \bibinfo{pages}{65--73}.
\newblock


\bibitem[\protect\citeauthoryear{M{\"o}ller and Trumbore}{M{\"o}ller and
  Trumbore}{1997}]%
        {moller1997fast}
\bibfield{author}{\bibinfo{person}{Tomas M{\"o}ller} {and} \bibinfo{person}{Ben
  Trumbore}.} \bibinfo{year}{1997}\natexlab{}.
\newblock \showarticletitle{Fast, minimum storage ray-triangle intersection}.
\newblock \bibinfo{journal}{\emph{Journal of Graphics Tools}}
  \bibinfo{volume}{2}, \bibinfo{number}{1} (\bibinfo{year}{1997}),
  \bibinfo{pages}{21--28}.
\newblock


\bibitem[\protect\citeauthoryear{Pottmann, Asperl, Hofer, and Kilian}{Pottmann
  et~al\mbox{.}}{2008}]%
        {pottmann2008architectural}
\bibfield{author}{\bibinfo{person}{Helmut Pottmann}, \bibinfo{person}{Andreas
  Asperl}, \bibinfo{person}{Michael Hofer}, {and} \bibinfo{person}{Axel
  Kilian}.} \bibinfo{year}{2008}\natexlab{}.
\newblock \bibinfo{booktitle}{\emph{Architectural geometry}}.
\newblock \bibinfo{publisher}{Bentley Institute Press}.
\newblock


\bibitem[\protect\citeauthoryear{Pottmann, Eigensatz, Vaxman, and
  Wallner}{Pottmann et~al\mbox{.}}{2015}]%
        {pottmann2015architectural}
\bibfield{author}{\bibinfo{person}{Helmut Pottmann}, \bibinfo{person}{Michael
  Eigensatz}, \bibinfo{person}{Amir Vaxman}, {and} \bibinfo{person}{Johannes
  Wallner}.} \bibinfo{year}{2015}\natexlab{}.
\newblock \showarticletitle{Architectural geometry}.
\newblock \bibinfo{journal}{\emph{Computers \& graphics}}  \bibinfo{volume}{47}
  (\bibinfo{year}{2015}), \bibinfo{pages}{145--164}.
\newblock


\bibitem[\protect\citeauthoryear{Qu, Zhang, Gao, Jiang, and Chen}{Qu
  et~al\mbox{.}}{2019}]%
        {qu2019efficient}
\bibfield{author}{\bibinfo{person}{Ziyin Qu}, \bibinfo{person}{Xinxin Zhang},
  \bibinfo{person}{Ming Gao}, \bibinfo{person}{Chenfanfu Jiang}, {and}
  \bibinfo{person}{Baoquan Chen}.} \bibinfo{year}{2019}\natexlab{}.
\newblock \showarticletitle{Efficient and conservative fluids using
  bidirectional mapping}.
\newblock \bibinfo{journal}{\emph{ACM Transactions on Graphics (TOG)}}
  \bibinfo{volume}{38}, \bibinfo{number}{4} (\bibinfo{year}{2019}),
  \bibinfo{pages}{1--12}.
\newblock


\bibitem[\protect\citeauthoryear{Ramponi, Blocken, Laura, and Janssen}{Ramponi
  et~al\mbox{.}}{2015}]%
        {RN242}
\bibfield{author}{\bibinfo{person}{Rubina Ramponi}, \bibinfo{person}{Bert
  Blocken}, \bibinfo{person}{B Laura}, {and} \bibinfo{person}{Wendy~D
  Janssen}.} \bibinfo{year}{2015}\natexlab{}.
\newblock \showarticletitle{CFD simulation of outdoor ventilation of generic
  urban configurations with different urban densities and equal and unequal
  street widths}.
\newblock \bibinfo{journal}{\emph{Building and Environment}}
  \bibinfo{volume}{92} (\bibinfo{year}{2015}), \bibinfo{pages}{152--166}.
\newblock
\showISSN{0360-1323}


\bibitem[\protect\citeauthoryear{Rando, Mu\~{n}oz, and Patow}{Rando
  et~al\mbox{.}}{2016}]%
        {10.5555/3056883.3056894}
\bibfield{author}{\bibinfo{person}{Eduard Rando}, \bibinfo{person}{Imanol
  Mu\~{n}oz}, {and} \bibinfo{person}{Gustavo Patow}.}
  \bibinfo{year}{2016}\natexlab{}.
\newblock \showarticletitle{Interactive Low-Cost Wind Simulation for Cities}.
  In \bibinfo{booktitle}{\emph{Proceedings of the Eurographics Workshop on
  Urban Data Modelling and Visualisation}} (Li\`{e}ge, Belgium)
  \emph{(\bibinfo{series}{UDMV '16})}. \bibinfo{publisher}{Eurographics
  Association}, \bibinfo{address}{Goslar, DEU}, \bibinfo{pages}{57–63}.
\newblock
\showISBNx{9783038680130}


\bibitem[\protect\citeauthoryear{Saad}{Saad}{2003}]%
        {saad2003iterative}
\bibfield{author}{\bibinfo{person}{Yousef Saad}.}
  \bibinfo{year}{2003}\natexlab{}.
\newblock \bibinfo{booktitle}{\emph{Iterative methods for sparse linear
  systems}}.
\newblock \bibinfo{publisher}{SIAM}.
\newblock


\bibitem[\protect\citeauthoryear{Selle, Fedkiw, Kim, Liu, and Rossignac}{Selle
  et~al\mbox{.}}{2008}]%
        {selle2008unconditionally}
\bibfield{author}{\bibinfo{person}{Andrew Selle}, \bibinfo{person}{Ronald
  Fedkiw}, \bibinfo{person}{Byungmoon Kim}, \bibinfo{person}{Yingjie Liu},
  {and} \bibinfo{person}{Jarek Rossignac}.} \bibinfo{year}{2008}\natexlab{}.
\newblock \showarticletitle{An unconditionally stable MacCormack method}.
\newblock \bibinfo{journal}{\emph{Journal of Scientific Computing}}
  \bibinfo{volume}{35}, \bibinfo{number}{2} (\bibinfo{year}{2008}),
  \bibinfo{pages}{350--371}.
\newblock


\bibitem[\protect\citeauthoryear{Selle, Rasmussen, and Fedkiw}{Selle
  et~al\mbox{.}}{2005}]%
        {selle2005vortex}
\bibfield{author}{\bibinfo{person}{Andrew Selle}, \bibinfo{person}{Nick
  Rasmussen}, {and} \bibinfo{person}{Ronald Fedkiw}.}
  \bibinfo{year}{2005}\natexlab{}.
\newblock \showarticletitle{A vortex particle method for smoke, water and
  explosions}.
\newblock In \bibinfo{booktitle}{\emph{ACM SIGGRAPH 2005 Papers}}.
  \bibinfo{pages}{910--914}.
\newblock


\bibitem[\protect\citeauthoryear{Sharma, Mittal, and Gairola}{Sharma
  et~al\mbox{.}}{2018}]%
        {RN239}
\bibfield{author}{\bibinfo{person}{Ashutosh Sharma}, \bibinfo{person}{Hemant
  Mittal}, {and} \bibinfo{person}{Ajay Gairola}.}
  \bibinfo{year}{2018}\natexlab{}.
\newblock \showarticletitle{Mitigation of wind load on tall buildings through
  aerodynamic modifications}.
\newblock \bibinfo{journal}{\emph{Journal of Building Engineering}}
  \bibinfo{volume}{18} (\bibinfo{year}{2018}), \bibinfo{pages}{180--194}.
\newblock
\showISSN{2352-7102}


\bibitem[\protect\citeauthoryear{Stam}{Stam}{1999}]%
        {stam1999stable}
\bibfield{author}{\bibinfo{person}{Jos Stam}.} \bibinfo{year}{1999}\natexlab{}.
\newblock \showarticletitle{Stable fluids}. In
  \bibinfo{booktitle}{\emph{Proceedings of the 26th annual conference on
  Computer graphics and interactive techniques}}. \bibinfo{pages}{121--128}.
\newblock


\bibitem[\protect\citeauthoryear{Thordal, Bennetsen, and Koss}{Thordal
  et~al\mbox{.}}{2019}]%
        {RN238}
\bibfield{author}{\bibinfo{person}{Marie~Skytte Thordal},
  \bibinfo{person}{Jens~Chr Bennetsen}, {and} \bibinfo{person}{H~Holger~H
  Koss}.} \bibinfo{year}{2019}\natexlab{}.
\newblock \showarticletitle{Review for practical application of CFD for the
  determination of wind load on high-rise buildings}.
\newblock \bibinfo{journal}{\emph{Journal of Wind Engineering and Industrial
  Aerodynamics}}  \bibinfo{volume}{186} (\bibinfo{year}{2019}),
  \bibinfo{pages}{155--168}.
\newblock
\showISSN{0167-6105}


\bibitem[\protect\citeauthoryear{Wilcox}{Wilcox}{2006}]%
        {wilcox2006turbulence}
\bibfield{author}{\bibinfo{person}{David~C Wilcox}.}
  \bibinfo{year}{2006}\natexlab{}.
\newblock \bibinfo{booktitle}{\emph{Turbulence Modelling for CFD, 3rd
  edition}}.
\newblock \bibinfo{publisher}{DCW industries La Canada}, \bibinfo{address}{CA}.
\newblock


\bibitem[\protect\citeauthoryear{Zehnder, Narain, and Thomaszewski}{Zehnder
  et~al\mbox{.}}{2018}]%
        {zehnder2018advection}
\bibfield{author}{\bibinfo{person}{Jonas Zehnder}, \bibinfo{person}{Rahul
  Narain}, {and} \bibinfo{person}{Bernhard Thomaszewski}.}
  \bibinfo{year}{2018}\natexlab{}.
\newblock \showarticletitle{An advection-reflection solver for
  detail-preserving fluid simulation}.
\newblock \bibinfo{journal}{\emph{ACM Transactions on Graphics (TOG)}}
  \bibinfo{volume}{37}, \bibinfo{number}{4} (\bibinfo{year}{2018}),
  \bibinfo{pages}{1--8}.
\newblock


\bibitem[\protect\citeauthoryear{Zhang, Bridson, and Greif}{Zhang
  et~al\mbox{.}}{2015}]%
        {zhang2015restoring}
\bibfield{author}{\bibinfo{person}{Xinxin Zhang}, \bibinfo{person}{Robert
  Bridson}, {and} \bibinfo{person}{Chen Greif}.}
  \bibinfo{year}{2015}\natexlab{}.
\newblock \showarticletitle{Restoring the missing vorticity in
  advection-projection fluid solvers}.
\newblock \bibinfo{journal}{\emph{ACM Transactions on Graphics (TOG)}}
  \bibinfo{volume}{34}, \bibinfo{number}{4} (\bibinfo{year}{2015}),
  \bibinfo{pages}{1--8}.
\newblock


\bibitem[\protect\citeauthoryear{Zhongming, Linong, Wangqiang, and
  Wei}{Zhongming et~al\mbox{.}}{2021}]%
        {RN229}
\bibfield{author}{\bibinfo{person}{Zhu Zhongming}, \bibinfo{person}{Lu Linong},
  \bibinfo{person}{Zhang Wangqiang}, {and} \bibinfo{person}{Liu Wei}.}
  \bibinfo{year}{2021}\natexlab{}.
\newblock \showarticletitle{AR6 Climate Change 2021: The Physical Science
  Basis}.
\newblock   \bibinfo{volume}{IPCC Sixth Assessment Report (AR6)}
  (\bibinfo{year}{2021}).
\newblock


\end{thebibliography}

\section*{Acknowledgements}
This work has been supported by the KAUST Visual Computing Center and through individual baseline funding. The valuable comments of the anonymous reviewers that improved the manuscript are gratefully acknowledged.

\newpage

\begin{appendix}
\title{APPENDIX}
\section{Experiments and Validations}
\noindent Validations of our fluid simulator including its porosity model and the choice of the AI preconditioner are presented in this supplemental material.

\subsection{Validation of the Fluid Simulator}
\begin{figure}[htb]
    \centering
    \includegraphics[width=0.25\columnwidth]{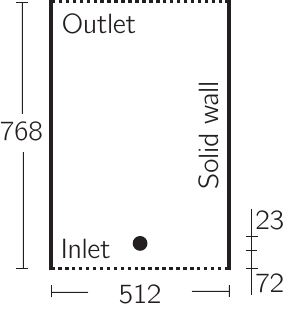}
    \includegraphics[width=0.475\columnwidth]{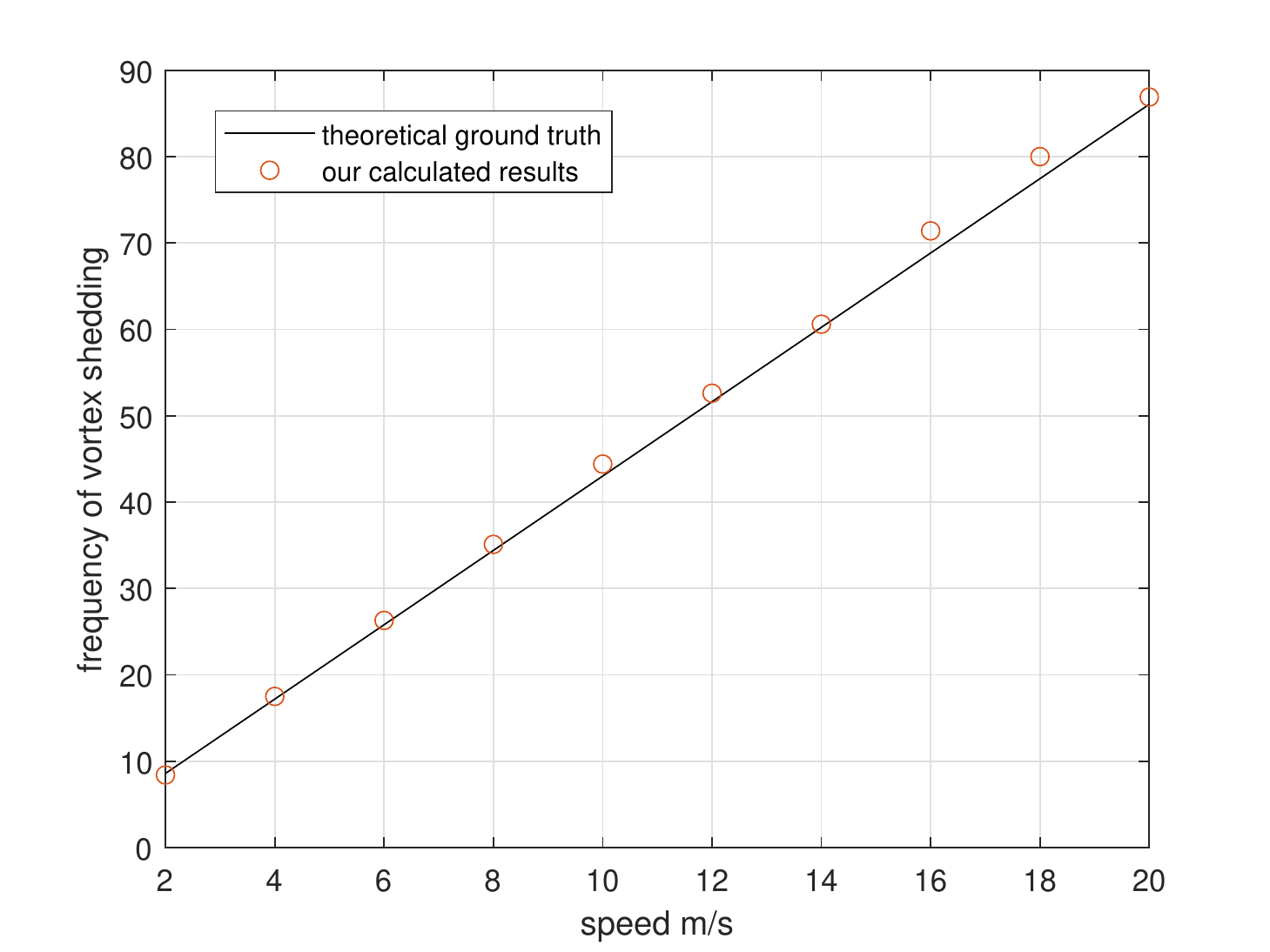}
    \caption{K\'arm\'an vortex street benchmark test case, frequency of vortex shedding vs. wind speed, geometric and simulation configuration(left).}
    \label{fig:solver-accuracy-test}
\end{figure}

As a standard benchmark example, we use a K\'arm\'an vortex street example simulating the stream around a cylinder to validate the accuracy of our solver, and the following widely-used analytical formula\footnote{https://thermopedia.com/content/1247/} (valid for $250 < Re < 2.5\cdot 10^5$ ) is used as the ground truth:
\begin{equation}
\centering
    St=0.198 \, (1-19.7/Re)\,,~~~~~~ f = St \, \cdot U / L\,,
\end{equation}
in which $St$ denotes the Strouhal number, $Re$ denotes the Reynolds number, and $U$ denotes wind speed, $f$ is the frequency of the vortex shedding, $L$ is the characteristic length and equal to the diameter of the cylinder. The calculated results are shown in the Figure \ref{fig:solver-accuracy-test}. For this simulation, the bottom boundary is set to be inlet, the top boundary is outlet, and both side boundaries are solid walls. The grid resolution is $512 \times 768$, $\Delta x = \Delta y = 2\cdot 10^{-3}$~m, $\Delta t = 1 \cdot 10^{-4}$ ~s, the diameter of the circular obstacle is 0.046~m. The inlet wind speed varies from 2~m/s to 20~m/s, thus the corresponding Reynolds number ranges from $5.86 \cdot 10^3$ to $5.86 \cdot 10^4$. As shown in Figure \ref{fig:solver-accuracy-test} our numerical results match the theoretical prediction very well. 

\subsection{Validation of the Porosity Model}
According to its classical definition, porosity can be quantified by a parameter $\phi= 1- V_s/V$, where $V$ is the total volume and the $V_s$ is the non-empty volume. Please note, that this definition assumes that gas communicates between the pores and the surrounding volume. In practice, this means that the pores must not be closed cavities.

Trees are canonical examples for porous objects. A common model for the computation of the drag force for tree objects is given by $\textbf{f}_d = -\rho_{air} C_d \text{LAD} |\textbf{u}| \textbf{u}$ with leaf area density (LAD). For buildings, LAD can, e.g., be replaced by $a((1-\phi)/\phi)^b$ with $a=0.62$ and $b= 2.5$. A drag force coefficient of $C_d = 1.0$ corresponds to a smooth surface.

\begin{figure}[htb]
    \centering
     \includegraphics[width=0.25\columnwidth]{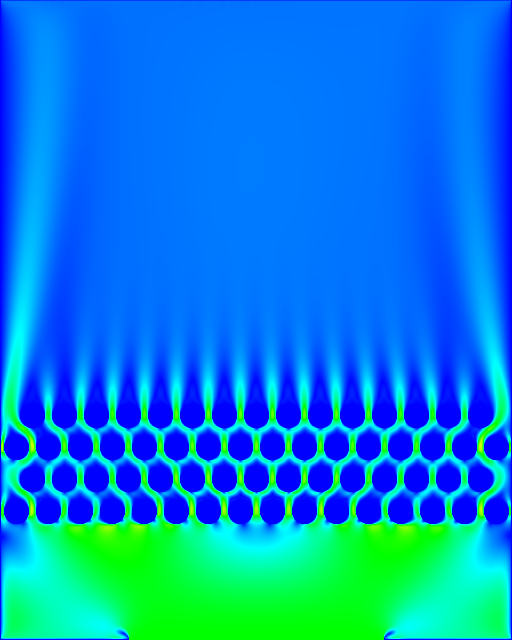}
    \includegraphics[width=0.55\columnwidth]{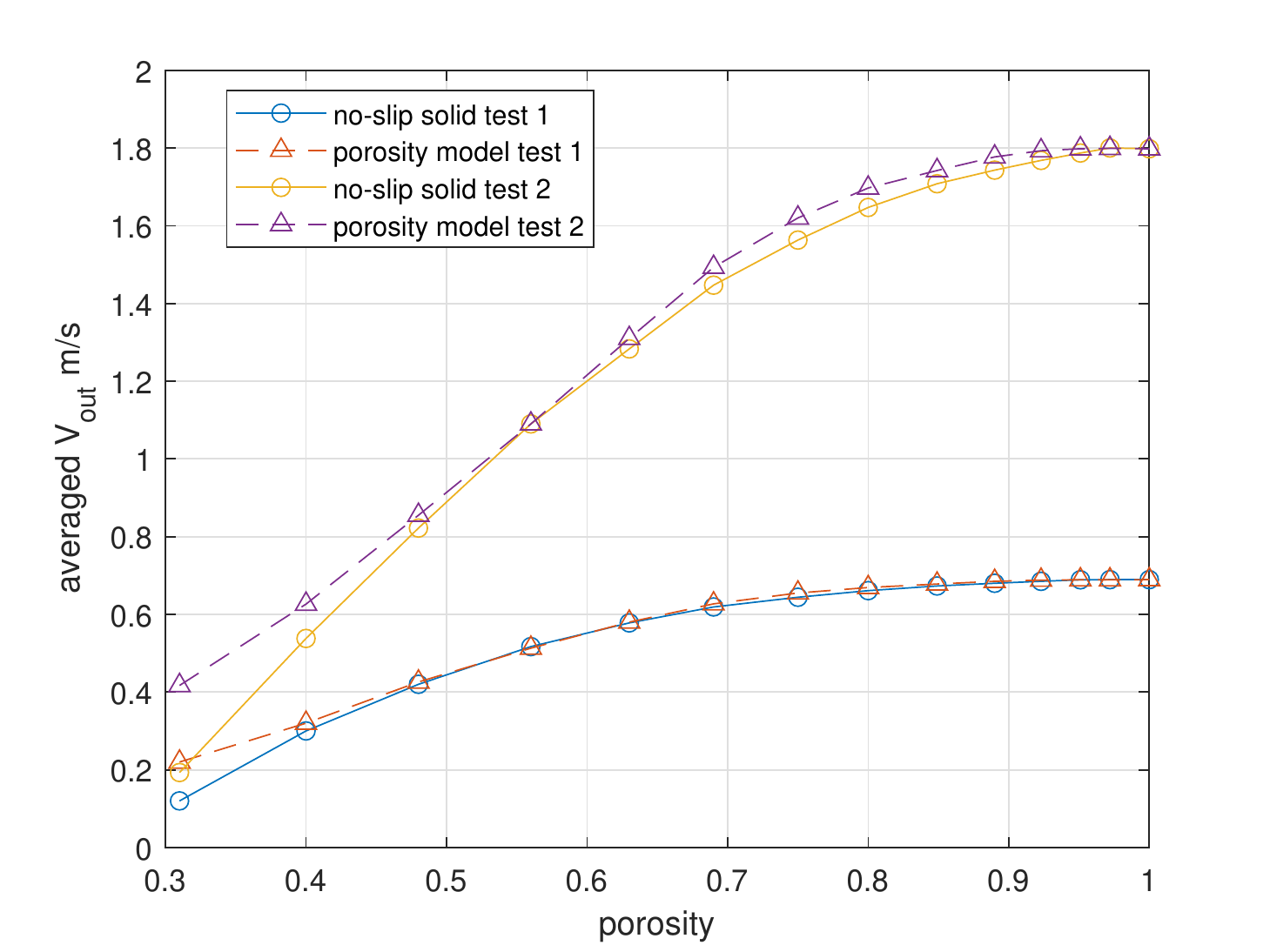}
    \caption{Benchmark test validating the porosity model.}
    \label{fig:porosit-test}
\end{figure}

We set up a validation experiment shown in Figure \ref{fig:porosit-test}. Form bottom to top, wind a blowing with a speed of 2~m/s and 5~m/s for Test 1 and Test 2 respectively. only the middle part of the bottom boundary is set to be inlet, its left and right parts are set to be outlet. Both, the side boundaries are set to be no-slip solid boundary. $V_\text{out}$ measures the averaged wind speed of the top outlet boundary. The different porosity configurations are done by changing the radius of the circular obstacles. The left sub-figure of Figure~\ref{fig:porosit-test} shows the dependence of $V_\text{out}$ on the porosity. In this simulation, a grid resolution of $512\times640$, $\Delta x = \Delta y = 1$~m, and $\Delta t = 0.1$~s are used. The results obtained using our porosity model are validated against the results calculated by using no-slip boundary conditions which are treated as the ground truth. It can clearly be observed that our porosity matched the ground truth very well.

\subsection{Efficiency of the AI Preconditioner}
\begin{figure}
    \centering
    \includegraphics[width=0.275\columnwidth]{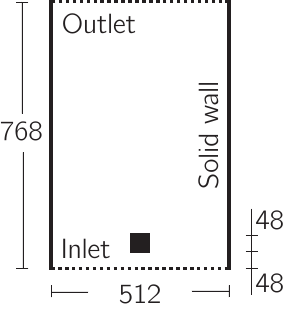}
    \includegraphics[width=0.55\columnwidth]{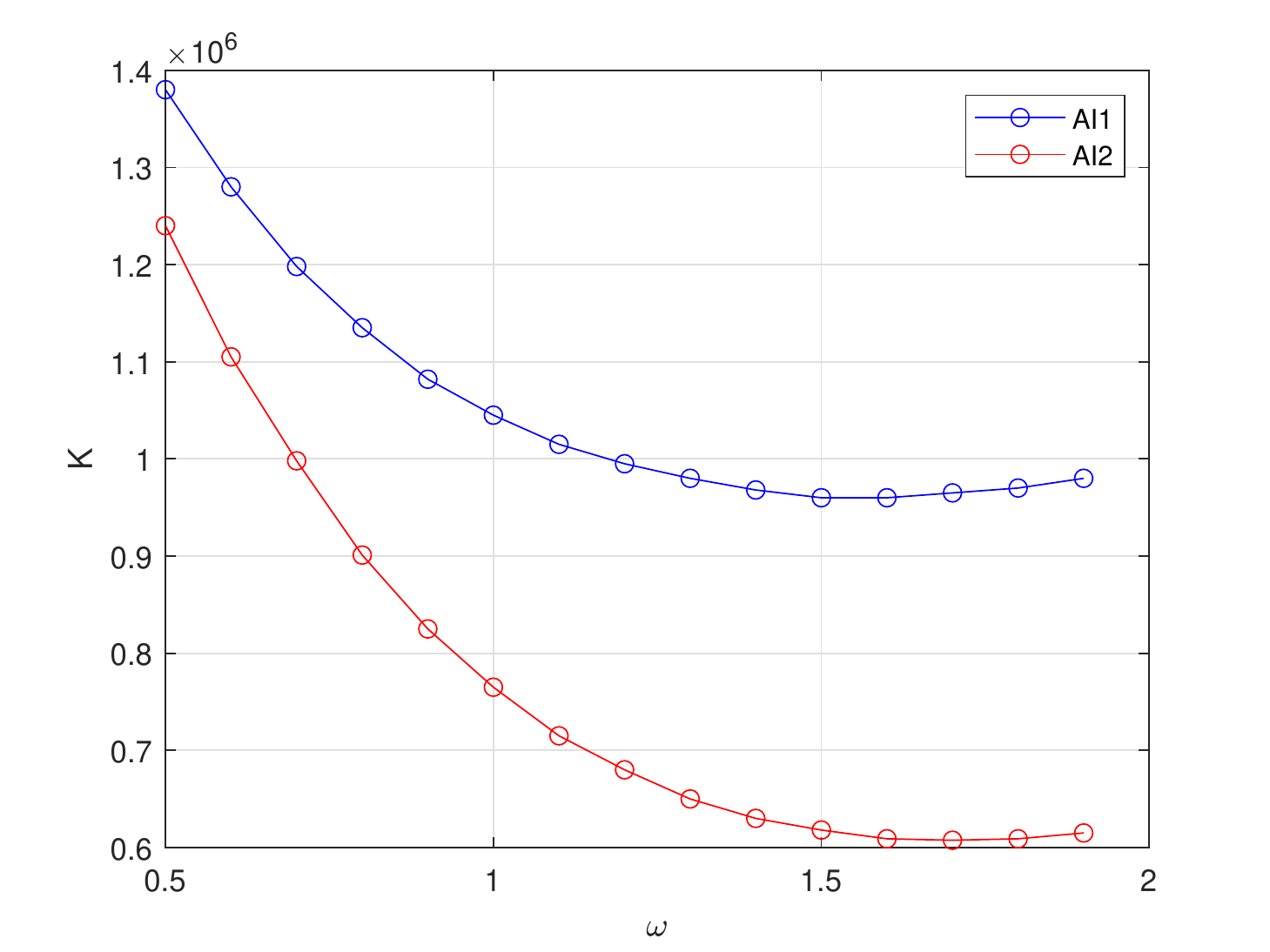}
    \caption{Illustration of the dependence of the condition number on $\omega$ using the approximate inverse preconditioner.}
    \label{fig:cond-over-w}
\end{figure}

We compare the AI preconditioner against popular preconditioners including Jacobi, SSOR, incomplete Cholesky factorization (IC) and modified incomplete Cholesky factorization (MIC). Since all these preconditioners can be precomputed for our simulation, we only evaluate their application process.
The AI preconditioner only requires one matrix-vector multiplication in its application process as the Jacobi preconditioner in each iteration, and is straightforward to parallelize. It is well known that the popular MIC preconditioner cannot be easily parallelized because of its inherently serial back substitutions. Therefore, we will mainly compare their convergence rate via measuring the condition number of the preconditioned matrix $\textbf{M}^{-1}\textbf{A}$ as shown in Table~\ref{tab:cond_precond},
\begin{table}[htb]
\small
    \caption{Comparison of different CG preconditioners.}
    \centering
    \begin{tabular}{c c c c c }
    \hline
         &  CG & Jacobi & AI1 & AI2 \\
         \hline
    $K$ & 2.3501e+06  & 2.3463e+06 & 9.5582e+05 & 6.0811e+05 \\
    \hline
    \end{tabular}
    \label{tab:cond_precond}
\end{table}
in which AI1 and AI2 denote the AI preconditioners using first and second order approximation, and their condition number depends on the value of $\omega$ as show in Figure~\ref{fig:cond-over-w}. The optimal value of $\omega$ is between 1.6 and 1.7.

The calculation of the condition numbers for SSOR, IC, and MIC involves the inverse matrix calculation of a large scale matrix. It is too time and too memory expensive to be calculated in our workstation for the current resolution. Thus, we perform another sparser simulation ($128 \times 192$) where SSOR, IC, and MIC are included in the comparison of the condition number shown in Table~\ref{tab:cond_precond2}. 
\begin{table}[htb]
\small
\caption{Comparison of different CG preconditioners using a sparser grid resolution ($128 \times 192$).}
\centering
\begin{tabular}{c c c c c }
\hline
&  CG & Jacobi & AI1 & AI2   \\
\hline
$K$ & $1.4516\cdot 10^5$ & $1.4422\cdot 10^5$ & $5.8736\cdot 10^4$ & $3.7385\cdot 10^4$ \\
\hline
& SSOR & IC & MIC &  \\
\hline
$K$ & $1.2234\cdot 10^4$ & $2.1118\cdot 10^4$ & $4.7955\cdot 10^3$ &  \\
\hline
\end{tabular}
\label{tab:cond_precond2}
\end{table}
 
The condition number of the matrix $\textbf{M}^{-1}\textbf{A}$ is defined as the ratio of its maximum and minimum eigenvalues. The lower condition number of an arbitrary symmetric positive definite matrix is, the faster the CG algorithm converges. Although the AI preconditioner does not have the best convergence rate, it has the best balance between convergence rate and the parallel computational cost of each iteration.
We employ the converge condition
\begin{equation}
{(\mathbf{M^{-1}r}) \cdot \mathbf{r}}/{\|\mathbf{b}\|^2} < 10^{-5}\,,
\end{equation} 
in which $\mathbf{r}$ denotes the residual. CG without any preconditioner converged after 350 iterations which took 0.62~s. Using the Jacobi preconditioner, CG converged after 330 iterations which took 0.55~s. Using AI1, only 5 iterations are required for convergence which took 0.01~s. AI1 and AI2 show comparable performance. In these times comparisons, we do not include other preconditioners such as SSOR because they need backward substitutions which is suboptimal for the parallelization \cite{chu2017schur}.

\end{appendix}

\end{document}